\documentclass[a4paper,onecolumn,11pt,unpublished]{quantumarticle}
\pdfoutput=1
\usepackage[utf8]{inputenc}
\usepackage[english]{babel}
\usepackage[T1]{fontenc}
\usepackage{amsmath,amssymb,amsthm,mathtools}
\usepackage{booktabs}
\usepackage{adjustbox}
\usepackage{graphicx}
\usepackage{tikz}
\usetikzlibrary{arrows.meta,positioning,shapes.geometric,calc,decorations.pathreplacing}
\usepackage{algorithm}
\usepackage{algpseudocode}
\usepackage[numbers,sort&compress]{natbib}
\usepackage{microtype}
\usepackage{hyperref}
\usepackage[capitalise]{cleveref}
\crefname{figure}{Figure}{Figures}
\Crefname{figure}{Figure}{Figures}

\newtheorem{theorem}{Theorem}
\newtheorem{lemma}[theorem]{Lemma}
\newtheorem{proposition}[theorem]{Proposition}
\newtheorem{corollary}[theorem]{Corollary}
\newtheorem{conjecture}{Conjecture}
\theoremstyle{definition}
\newtheorem{definition}[theorem]{Definition}
\theoremstyle{remark}
\newtheorem{remark}[theorem]{Remark}

\newcommand{\E}{E}
\newcommand{\UB}{\mathrm{UB}}
\newcommand{\LB}{\mathrm{LB}}
\newcommand{\Hnode}{\hat H}
\newcommand{\lex}{<_{\mathrm{lex}}}
\newcommand{\comp}{\mathrm{comp}}
\newcommand{\lo}{\mathrm{lo}}
\newcommand{\hi}{\mathrm{hi}}
\DeclareMathOperator{\lexmin}{lexmin}

\begin{document}

\title{Proven optimality for LABS with quantum-classical branch and bound}
\author{Andr\'as Cz\'egel}
\affiliation{Bentax Plus, Budapest, Hungary}
\orcid{0009-0007-9156-0124}
\maketitle

\begin{abstract}
Low autocorrelation binary sequences (LABS) are the phase codes of pulse compression radar and
the synchronization words of digital communication, and finding the sequence of a given length
whose aperiodic autocorrelations have the smallest sum of squares is a hard binary optimization
problem. The quantum optimization methods tested on LABS so far return good
sequences, but they cannot show that a sequence is optimal or bound how far from optimal it may
be. We show that a quantum optimization algorithm can deliver proven optimality. In our
quantum-classical branch and bound, fixed-angle QAOA samples the Hamiltonians of the nodes of
the search tree, its samples supply candidate sequences and the branching choice, and classical
bounds prune the tree, so that the algorithm returns either a proof of optimality or, if it is
stopped early, the best sequence found with a certified lower bound and gap. It works on the
native quartic Hamiltonian without penalty terms, and its correctness does not depend on the
sampler. With QAOA at depth $p=12$ and at $p=1$ and $1{,}000$ shots per node, it proved the
optima of all lengths from $2$ to $30$, where from length $27$ on the root node is split without
being sampled. Every run records both bounds against time, and its primal-dual integral measures its
convergence. We claim no speedup.
\end{abstract}

\section{Introduction}
\label{sec:intro}

A binary sequence $s\in\{-1,+1\}^n$ has the aperiodic autocorrelations
\begin{equation*}
  C_k(s)=\sum_{i=1}^{n-k}s_is_{i+k},\qquad k=0,\dots,n-1 .
\end{equation*}
The value $C_0=n$ is the peak; the $C_k$ with $k\ge1$ are the sidelobes. The low autocorrelation
binary sequence (LABS) problem asks for a sequence of given length with the smallest energy
$E(s)=\sum_{k=1}^{n-1}C_k(s)^2$, the sum of the squared sidelobes, or equivalently with the largest merit factor
$\mathcal{F}=n^2/(2E)$ \citep{golay1972class} (written $F$ in the literature; here $F$ is
reserved for the number of free spins of a node).

Such sequences are used wherever a known signal is located in time by correlation. In pulse
compression radar and sonar a long pulse is phase-coded by $s$, the matched filter outputs the
aperiodic autocorrelation of $s$, and the sidelobes appear as false echoes of relative height
$|C_k|/n$ \citep{boehmer1967binary,levanon2004radar,golomb2005signal}. The same sequences serve as
spreading sequences \citep{sarwate1980crosscorrelation}, synchronization words
\citep{massey1972optimum} and probe sequences for channel estimation \citep{tellambura1998channel}.
In analysis, LABS is Littlewood's question of how flat a polynomial with coefficients $\pm1$ can be
on the unit circle \citep{jedwab2013littlewood}, and in statistical physics it is the Bernasconi
model \citep{bernasconi1987low}, which is frustrated and glassy although its couplings contain no
randomness, with isolated ground states among many local minima
\citep{bernasconi1987low,bouchaud1994self,marinari1994replica,ferreira2000landscape}, so that
local search is hard. Since each length $n$ defines a single instance, progress is easy to
measure, and LABS has become a standard target \citep{shaydulin2024qaoa} for the quantum
approximate optimization algorithm (QAOA) \citep{farhi2014qaoa}.

Several quantum optimization methods have been tested on LABS: fixed-angle QAOA in noiseless
simulation up to $40$ qubits and on trapped-ion devices up to $18$ qubits
\citep{shaydulin2024qaoa}, bias-field digitized counterdiabatic quantum optimization (BF-DCQO) in
simulation up to $n=30$ and on an IBM device at $n=20$ \citep{koch2026qoblib}, digitized
counterdiabatic quantum optimization (DCQO) in simulation as a seed for memetic tabu search up to
$n=37$ \citep{gomezcadavid2026scaling}, and a sampled guided quantum walk in simulation up to $30$
qubits \citep{nzongani2026sampled}. All of them measure how fast, or how closely, a heuristic
reaches an optimum that is already known. None proves that a sample is optimal or bounds its gap
when the optimum is unknown, and none of these works reports a certified bound against time, so neither two such methods nor a quantum and a classical
solver can be compared by a proof or a certified gap.

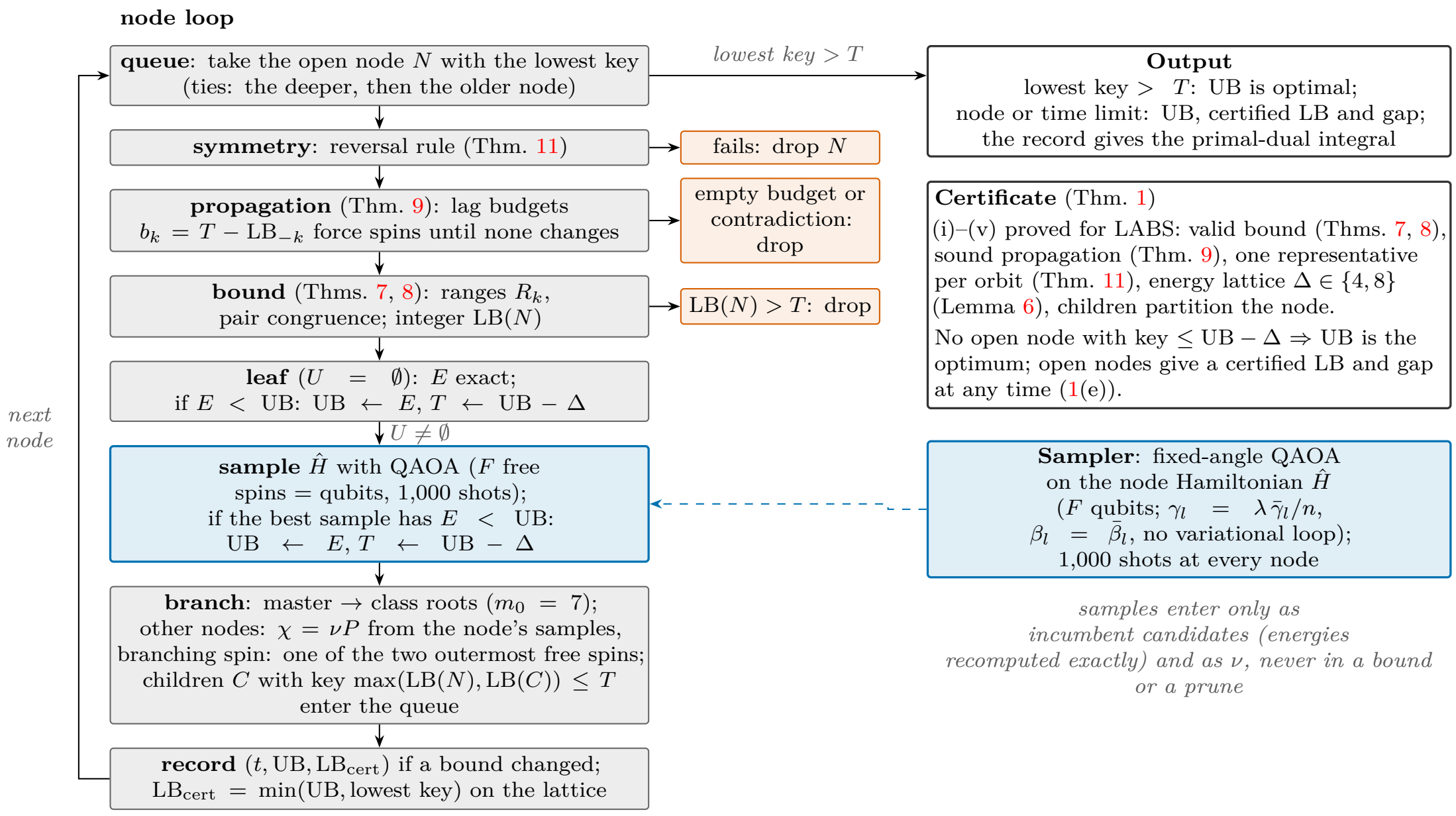
\begin{figure}[tbp]
\centering
\resizebox{\textwidth}{!}{
\definecolor{qblue}{HTML}{0072B2}
\definecolor{qverm}{HTML}{D55E00}
\begin{tikzpicture}[
  x=1mm, y=1mm, font=\scriptsize,
  box/.style={draw, rounded corners=1.2pt, align=center, inner sep=2.5pt, line width=0.5pt},
  stepbox/.style={text width=68mm},
  quantum/.style={box, draw=qblue, fill=qblue!12, line width=0.8pt},
  classical/.style={box, draw=black!60, fill=black!7},
  prune/.style={box, draw=qverm, fill=qverm!10, text width=24mm},
  cert/.style={box, draw=black!80, fill=white, line width=0.8pt},
  note/.style={align=center, font=\scriptsize\itshape, text=black!65},
  arr/.style={-{Stealth[length=1.6mm]}, line width=0.5pt},
  dasharr/.style={-{Stealth[length=1.6mm]}, line width=0.5pt, qblue, dashed},
]
  \hyphenpenalty=10000 \exhyphenpenalty=10000
  \node[classical, stepbox] (q) at (0,0)
    {\textbf{queue}: take the open node $N$ with the lowest key\\ (ties: the deeper, then the older node)};
  \node[classical, stepbox, below=3mm of q] (sym)
    {\textbf{symmetry}: reversal rule (Thm.~\labelcref{thm:sym})};
  \node[classical, stepbox, below=3mm of sym] (prop)
    {\textbf{propagation} (Thm.~\labelcref{thm:prop}): lag budgets\\ $b_k=T-\LB_{-k}$ force spins until none changes};
  \node[classical, stepbox, below=3mm of prop] (bnd)
    {\textbf{bound} (Thms.~\labelcref{thm:perlag}, \labelcref{thm:pair}): ranges $R_k$,\\ pair congruence; integer $\LB(N)$};
  \node[classical, stepbox, below=3mm of bnd] (leaf)
    {\textbf{leaf} ($U=\emptyset$): $\E$ exact;\\ if $\E<\UB$: $\UB\gets\E$, $T\gets\UB-\Delta$};
  \node[quantum, stepbox, below=3mm of leaf] (smp)
    {\textbf{sample} $\Hnode$ with QAOA ($F$ free spins $=$ qubits, $1{,}000$ shots);\\ if the best sample has $\E<\UB$: $\UB\gets\E$, $T\gets\UB-\Delta$};
  \node[classical, stepbox, below=3mm of smp] (br)
    {\textbf{branch}: master $\to$ class roots ($m_0=7$);\\
     other nodes: $\chi=\nu P$ from the node's samples,\\
     branching spin: one of the two outermost free spins;\\
     children $C$ with key $\max(\LB(N),\LB(C))\le T$\\ enter the queue};
  \node[classical, stepbox, below=3mm of br] (rec)
    {\textbf{record} $(t,\UB,\LB_{\mathrm{cert}})$ if a bound changed;\\
     $\LB_{\mathrm{cert}}=\min(\UB,\text{lowest key})$ on the lattice};
  \draw[arr] (q) -- (sym); \draw[arr] (sym) -- (prop); \draw[arr] (prop) -- (bnd);
  \draw[arr] (bnd) -- (leaf); \draw[arr] (leaf) -- node[right, note] {$U\ne\emptyset$} (smp);
  \draw[arr] (smp) -- (br); \draw[arr] (br) -- (rec);
  \draw[arr] (rec.west) -- ++(-4,0) |- (q.west) node[pos=0.25, left, note, text width=10mm] {next node};
  \node[above=0.8mm of q.north west, anchor=south west, font=\scriptsize\bfseries] {node loop};

  \node[prune, right=4mm of sym] (pr0) {fails: drop $N$};
  \node[prune, right=4mm of prop] (pr1) {empty budget or\\ contradiction: drop};
  \node[prune, right=4mm of bnd] (pr2) {$\LB(N)>T$: drop};
  \draw[arr] (sym) -- (pr0); \draw[arr] (prop) -- (pr1); \draw[arr] (bnd) -- (pr2);

  \node[cert, text width=66mm, anchor=north west] (out) at ($(pr0.east |- q.north)+(6,0)$)
    {\textbf{Output}\\ lowest key $>T$: $\UB$ is optimal;\\
     node or time limit: $\UB$, certified $\LB$ and gap;\\
     the record gives the primal-dual integral};
  \draw[arr] (q.east) -- (q.east -| out.west) node[pos=0.5, above, note] {lowest key $>T$};
  \node[cert, text width=66mm, align=flush left, below=3mm of out] (cert)
    {\textbf{Certificate} (Thm.~\labelcref{thm:framework})\\[2pt]
     (i)--(v) proved for LABS: valid bound (Thms.~\labelcref{thm:perlag}, \labelcref{thm:pair}),
     sound propagation (Thm.~\labelcref{thm:prop}), one representative per orbit
     (Thm.~\labelcref{thm:sym}), energy lattice $\Delta\in\{4,8\}$ (Lemma~\labelcref{lem:lattice}),
     children partition the node.\\[2pt]
     No open node with key $\le\UB-\Delta$ $\Rightarrow$ $\UB$ is the optimum; open nodes give a
     certified $\LB$ and gap at any time (\labelcref{thm:framework}(e)).};
  \node[quantum, text width=66mm, anchor=north west] (sampler) at ($(cert.south west)+(0,-4)$)
    {\textbf{Sampler}: fixed-angle QAOA on the node Hamiltonian $\Hnode$\\
     ($F$ qubits; $\gamma_l=\lambda\,\bar\gamma_l/n$, $\beta_l=\bar\beta_l$, no variational loop);\\
     $1{,}000$ shots at every node};
  \draw[dasharr] (sampler.west) -- ++(-5,0) |- (smp.east);
  \node[note, text width=66mm, below=1.5mm of sampler] (snote)
    {samples enter only as incumbent candidates (energies\\ recomputed exactly) and as $\nu$, never in a bound\\
     or a prune};

\end{tikzpicture}}
\caption{QCBB for LABS. Left: the node loop. The open node with the lowest bound is taken from
the queue, propagated, bounded and sampled. The blue box calls the sampler, fixed-angle QAOA on
the node Hamiltonian $\Hnode$ with $F$ qubits, where $F$ is the number of free spins of the
node; gray boxes are classical, vermilion boxes are the pruning tests, and the output is outlined
in black. The samples enter only as incumbent candidates and as the weights $\nu$ of the
conflict values. After every node, $\UB$ and the certified $\LB$ are recorded with the time, and
the primal-dual integral is computed from this record. Right: the output, the certificate and the
sampler. Conditions (i)--(v) of \cref{thm:framework} are each proven for LABS in the result
named, and none refers to the samples.}
\label{fig:flow}
\end{figure}

This paper presents a quantum optimization algorithm that proves optimality and measures its
convergence. Fixed-angle QAOA on the Hamiltonian of each node is the sampler inside an exact
quantum-classical hybrid branch and bound, QCBB \citep{czegel2025qcbb} (\cref{fig:flow}). The
samples supply incumbent candidates (the best sequence found so far is the incumbent; its energy
is $\UB$) and the conflict values that select the branching variable, while classical bounds prune
the tree and certify optimality (\cref{thm:framework}). Convergence is measured by the
primal-dual integral (\cref{sec:setup}). The aim is not to beat classical solvers:
\citep{packebusch2016labs} proved the optima of every $n\le66$ with a branch and bound that spends
microseconds per node, whereas ours makes one QAOA call at every node it branches, the root at $n\ge27$ excepted. Our
contribution is a proof of these optima by a quantum optimization algorithm, with a certificate
and a measured convergence that the existing quantum approaches to LABS do not give.

Our contributions are the following.
\begin{itemize}
\item To our knowledge, the first optimality proofs for LABS by a quantum optimization
algorithm, and the first primal-dual integral reported for a quantum algorithm on LABS. With
fixed-angle QAOA as the sampler, the algorithm proved the optima of all lengths $n=2$--$30$ at
depth $p=12$ and at $p=1$ (\cref{sec:experiments}).
\item A statement of QCBB for a general binary problem with one correctness theorem: five
conditions on an instance under which the algorithm returns a proven optimum or a valid gap
(\cref{thm:framework}). \Cref{tab:components} lists what is reused from QCBB and what is new.
\item Proofs that LABS satisfies the five conditions on its native quartic Hamiltonian, without
linearization or penalties: an energy lattice, with a congruence modulo~8 when $n\equiv2\pmod 4$
(\cref{lem:lattice}); an integer bound from the exact set of values that each autocorrelation
can still take, coupled across lags modulo~4 (\cref{thm:perlag,thm:pair}); a propagation rule
(\cref{thm:prop}); and a symmetry reduction (\cref{thm:sym}).
\item An analysis of what the samples can change in a proof. On nodes whose fixed spins form a
prefix and a suffix, the conflict values reproduce the classical branching order of
\citep{mertens1996exhaustive,packebusch2016labs}, which fixes the spins from the two ends inward
(\cref{prop:frontier}). The samples change the search tree only through the time at which the
incumbent becomes optimal and through the choice between the two outermost free spins at each
branched node (\cref{prop:samples}).
\item A fixed-angle QAOA sampler for the Hamiltonians of the nodes, with the published fixed
angles of the master Hamiltonian rescaled (\cref{sec:sampler}).
\item Resource counts of the QAOA circuits for present devices, and emulations of one job with
the noise of an IBM device and of a Quantinuum device, with and without the Iceberg error
detection code \citep{he2025iceberg,jin2025iceberg} (\cref{sec:hardware}).
\end{itemize}

What is proved, measured, estimated and conjectured is the following. Proved: the correctness
theorem and its LABS instance (\cref{thm:framework,cor:labs}; \cref{app:framework,app:lattice,app:bounds,app:prop})
and the two propositions on conflict values (\cref{app:conflict}). Measured: the proofs of
$n=2$--$30$ with QAOA at $p=12$ and at $p=1$, their cost and their bounds against time
(\cref{sec:experiments}), what the samples contribute (\cref{sec:samples}) and the
noise emulations (\cref{sec:hardware}). Estimated: the gate budget of present devices
(\cref{sec:hardware}). Conjectured, from measurements only: the angle rescaling
(\cref{sec:discussion}); the growth of the search tree is reported as a fit with its interval
(\cref{sec:experiments}).

Little is known about the largest merit factor $\mathcal{F}_n$ beyond small $n$. A non-rigorous
argument estimates that $\mathcal{F}_n$ tends to about $12.32$ \citep{golay1982merit}, whereas
the best proven lower bound on the limit superior of the merit factor along a family of
sequences was $6$, for rotated Legendre sequences \citep{hoholdt1988legendre}, until it was
raised to $6.342061\ldots$ \citep{jedwab2013littlewood,jedwab2013advances}, and whether
$\mathcal{F}_n$ is bounded at all is open \citep{packebusch2016labs}. Exact values of
$\mathcal{F}_n$ are therefore known only for the lengths at which the LABS problem has been
solved exactly.

Exact methods examine the $2^n$ sequences, with or without pruning. Plain enumeration in Gray-code order costs
$\Theta(n2^n)$ time \citep{packebusch2016labs}, and branch and bound improves on it by discarding
partial sequences whose lower bound exceeds the energy of the best sequence found so far. The branch and bound of \citep{mertens1996exhaustive} has a runtime
growing like $1.85^n$ and proved the optima up to $n=48$, and \citep[Sec.~3]{prestwich2013labs}
reports $1.80^n$ in wall-clock time ($1.74^n$ in recursive calls). A branch and bound with a
bound based on the congruence $C_k+C_{n-k}\equiv n\pmod 4$, run on $18$ GPUs, proved the optima
up to $n=64$ with a runtime growing like $1.79^n$ \citep{wiggenbrock2010labs}, as reported in
\citep[Sec.~3.1.3]{packebusch2016labs}. Building on this, \citep{packebusch2016labs} tightened
the bounds and searched symmetry classes in parallel, using the best energy found by heuristics as
a fixed threshold; its runtime grows like $1.73^n$ as measured there and proved the optima
up to $n=66$. This is still the largest proven length, as no optimum beyond $n=66$ has been proved
\citep{koch2026qoblib,zhang2025new}. General-purpose solvers trail it: \citep{elloumi2021solving}
solve a quadratic reformulation by branch and bound, with LABS as one test set. In the
measurements of \citep{shaydulin2024qaoa}, the time of Gurobi to a certificate of optimality
grows by a fitted factor of $1.76$ per unit of $n$ ($95\%$ interval $1.72$--$1.79$), consistent
with $1.73$, but the largest instance solved to optimality in this way is $n=41$
\citep{koch2026qoblib}.

For odd $n=2m-1$, a sequence is skew-symmetric if $s_{m+l}=(-1)^ls_{m-l}$ for $l=1,\dots,m-1$.
Then $C_k=0$ for every odd $k$ \citep{golay1972class}, and only $2^m$ sequences remain, so the
restriction was proposed as a sieve for good sequences and conjectured to be asymptotically
optimal \citep{golay1977sieves,golay1982merit}. With the smaller space, branch and bound
restricted to skew-symmetric sequences has proved the skew-symmetric optima up to $n=89$ and then
$n=119$ \citep{prestwich2013labs,packebusch2016labs}. The restriction can miss the optimum, however: for
$10$ of the $31$ odd $n\le65$ no optimal sequence is skew-symmetric \citep{packebusch2016labs},
and \citep{zhang2025new} improved five odd lengths between $99$ and $119$ whose best known values
were optimal skew-symmetric sequences. A search restricted to skew-symmetric sequences therefore
gives an upper bound on the optimal energy $E^*$, not a proof of optimality over all sequences.

Heuristics find good sequences far beyond the proven range. \citep{bernasconi1987low} used
simulated annealing;
memetic tabu search \citep{gallardo2009memetic} is the best performing heuristic in the literature
for general LABS, with a time to reach a known optimum that grows like $1.34^n$
\citep{boskovic2017labs,shaydulin2024qaoa}, while self-avoiding walk solvers on skew-symmetric
sequences have the smallest runtime growth reported, $1.1504^n$ \citep{boskovic2017labs}. Both run in
parallel on a GPU \citep{boskovic2024parallel,zhang2025new}, and together they have found
sequences with $\mathcal{F}>9$ up to length $255$ \citep{brest2018heuristic,brest2022computational} and new
best known values for $16$ lengths between $92$ and $120$ \citep{zhang2025new}. These solvers
give no bound, so whether a best known value beyond $n=66$ is optimal is not known.

The quantum heuristics named above are measured in the same way. In the noiseless simulations of
\citep{shaydulin2024qaoa}, the time of fixed-angle QAOA to sample the optimum grows like $1.46^n$
at depth $p=12$, and like $1.21^n$ when QAOA is combined with quantum minimum finding, against
$1.34^n$ for memetic tabu search; on trapped-ion devices QAOA was executed with an
algorithm-specific error detection scheme up to $18$ qubits. BF-DCQO
\citep{gomezcadavid2025biasfield,romero2025biasfield} was applied to LABS with classical local
search on the samples \citep{koch2026qoblib}, and memetic tabu search seeded with DCQO samples has
a fitted time-to-solution growth of $1.24^n$ for $n=27$--$37$ \citep{gomezcadavid2026scaling}.
All of these growth rates are measured against optima that are known, the quantum ones on
$n\le40$.

A second line of work accelerates the tree search itself with quantum algorithms. Quantum walks
speed up backtracking near-quadratically \citep{montanaro2018backtracking} and estimate the size
of a tree \citep{ambainis2017tree}; \citep{montanaro2020branch,chakrabarti2022universal} accelerate branch and bound near-quadratically;
\citep{booth2021quantum} applies quantum backtracking in constraint programming; and
\citep{haner2024solving} designs a classical branch and bound with bounds cheap enough for such a
speedup. These works concern the cost of the tree search itself rather than near-term samplers.

A third line, the one closest to ours, places quantum samples inside a classical branch and bound.
\citep{peng2025hybrid} uses Ising solvers, among them a quantum annealer, as heuristics inside a
classical branch and bound and reports, against default Gurobi as the baseline, up to $17\%$
fewer nodes and $11\%$ less solution time, while \citep{simen2025branch} builds a branch-and-bound variant of BF-DCQO for
higher-order problems. In the bound-aware tree search of \citep{hadfield2026iqo}, measured
samples update a shared incumbent, valid classical bounds prune, and a complete search certifies
the optimum; on MaxCut the ordering guided by QAOA is non-random in certification tests up to
$n=28$. QCBB \citep{czegel2025qcbb} has the same division of labor: a branch and bound wraps an
arbitrary variational quantum sampler, whose samples give incumbents and conflict values for
branching, while a classical bound prunes the tree and certifies optimality, and the primal-dual
integral of a run is shown on a set partitioning instance. None of these works, however,
proves optimality on LABS or reports a primal-dual integral for a quantum algorithm on it.

QCBB was written for binary linear programs with equality constraints, with a quadratic Ising
Hamiltonian, big-$M$ penalties (a penalty constant $M$ that must exceed the objective range) and
a Goemans--Williamson bound, and none of these fits LABS: a product linearization of $\E$ needs
one binary for each pair product $s_is_j$, $n(n-1)/2$ of them, and one for each of its
$\lfloor n(n-2)(2n-5)/24\rfloor$ distinct quartic monomials, the sets $a<b<c<d$ with $a+d=b+c$
(each has the coefficient $4$, since it arises from the two lags $b-a$ and $c-a$), so about
$n^3/12$ binaries in all, and a penalty $M$ above the objective range, which is
$\max_s\E(s)=n(n-1)(2n-1)/6\sim n^3/3$ (the sum of $(n-k)^2$ over the lags, attained by the
constant sequence), far above the optimum $E^*\approx n^2/16$, while the
Goemans--Williamson slack (the gap between the semidefinite bound and the true minimum) exceeds
$E^*$ by orders of magnitude. We therefore state QCBB for a general binary problem, with a single
correctness theorem whose conditions name what an instance must supply, and build the LABS
instance on the native quartic Hamiltonian. What we keep from QCBB is the node loop, the single
loop that evaluates the nodes in turn and calls the sampler at each of them, so that the sampler
acts throughout the search and not only at its start. The quantum sampler is thus one component
of an exact algorithm, and any sampler enters the same comparison through the same metrics; for
instances beyond proven reach, the sampler's incumbent is the reported answer and sets the gap.

\section{Problem}
\label{sec:labs}

A sequence $s\in\{-1,+1\}^n$ has the aperiodic autocorrelations
$C_k(s)=\sum_{i=1}^{n-k}s_is_{i+k}$, $k=1,\dots,n-1$. The master problem is to find a sequence
that minimizes the energy
\begin{equation}
  \E(s)=\sum_{k=1}^{n-1} C_k(s)^2 .
  \label{eq:labs}
\end{equation}
We write $E^*$ for the optimal energy and read $s$ as a spin configuration ($\sigma$ in QCBB),
with $s_i$ the eigenvalue of the Pauli operator $Z_i$, so that the master Hamiltonian on $n$
qubits is the energy~\eqref{eq:labs} itself, $H(s)=\E(s)$. Expanding the squares gives $1$-,
$2$-, $3$- and $4$-body $Z$ terms with integer coefficients, although at the root there are no
linear or cubic terms. No linearization, big-$M$ or penalty is used, because $\E$ is already the
penalty form of the soft constraints $C_k=0$: whereas in QCBB the master Hamiltonian is built
from a binary linear program by a penalty transformation (QCBB, eqs.~(1)--(4)), here it is the
objective itself, so every eigenvalue is the energy of a sequence.

The energies do not take every integer value. Each $C_k$ has the parity of $n-k$, and
$\lfloor n/2\rfloor$ of the lags are odd, so every energy is congruent to $\lfloor n/2\rfloor$
modulo~$4$. For $n\equiv2\pmod 4$ more is true, because the congruence
$C_k+C_{n-k}\equiv n\pmod 4$ of \citep{wiggenbrock2010labs}, derived in
\citep[Sec.~3.1.3]{packebusch2016labs}, fixes, for every pair of even lags $k$
and $n-k$, the residue of $C_k^2+C_{n-k}^2$ modulo~$8$, so that all energies then lie in one
residue class modulo~$8$. Writing $\Delta=8$ for $n\equiv 2 \pmod 4$ and $\Delta=4$ otherwise, all
attainable energies lie on a lattice of step $\Delta$ (\cref{lem:lattice}, \cref{app:lattice}). So
if the incumbent value $\UB$ is attained and some sequence has $\E<\UB$, then $\E\le\UB-\Delta$,
and the search for a better sequence can skip the values that no sequence attains. The congruence
modulo~$8$ is not in \citep{packebusch2016labs,mertens1996exhaustive}, and we have not
found it elsewhere. It applies to one quarter of the instances, and its measured effect on the
search is small: with the optimum supplied at the root, the step $\Delta=8$ in place of $4$
leaves the node count unchanged at $n=22$ and $26$ and lowers it from $98{,}186$ to $91{,}534$
at $n=30$ (\cref{tab:ablation}).

\section{Quantum-classical hybrid branch and bound}
\label{sec:qcbb}

QCBB \citep{czegel2025qcbb} is a branch and bound in which the samples of a quantum sampler give
incumbents and guide branching, and classical bounds prune. Its definition requires five rules:
branching, node selection, bounding, pruning and stopping (QCBB Sec.~3). We describe them with
their LABS versions and refer to QCBB for everything that is unchanged. The rules are stated for a
general problem $\min\{f(x): x\in X\}$, $X=\{-1,+1\}^n$, with an integer-valued objective $f$; for
LABS, $f=\E$ and $x=s$.

A node $N=(\Phi,U)$ of the tree is a partial assignment: the spins in the fixed set $\Phi$ have
values, the spins in the free set $U$ have none, and $F=|U|$. The completions $\comp(N)\subseteq X$
of a node are the $x$ that agree with its fixed values. As in QCBB, a hat marks the node under
evaluation, and we call $f$ with the fixed values substituted, a function of the $F$ free spins,
the node Hamiltonian $\Hnode$; it is the reduced Hamiltonian of QCBB (eqs.~(10)--(16): column
elimination and the constant put aside), so its values are objective values of full solutions. The search works at a threshold. Because all objective
values lie on a lattice of step $\Delta$ (for LABS, \cref{sec:labs}; $\Delta=1$ for any integer
objective) and the incumbent value $\UB$ is attained, any $x$ with $f(x)<\UB$ has
$f(x)\le T:=\UB-\Delta$, and the search for an improvement is the search for completions with
$f\le T$. Before the first incumbent exists, $\UB=T=+\infty$, so no node is pruned by the bound
and the first leaf or sample sets $\UB$.

The optimality guarantee rests on five properties of the rules, made precise in
\cref{thm:framework} (\cref{app:framework}): (i) the node bound is valid; (ii) propagation removes
only completions with $f>T$; (iii) the symmetry rule keeps a representative of every orbit; (iv)
the objective values lie on a lattice of step $\Delta$; (v) the children of a node partition
its completions. The sampler enters none of them.

\Cref{alg:main} and \cref{fig:flow} summarize the LABS instance. The open nodes are kept in a
queue ordered by their lower bound, with the master node (the root, QCBB's master problem), which
has no spin fixed, as the first. Each step takes the open node with the lowest bound and evaluates
it: the symmetry rule (\cref{thm:sym}; for LABS the reversal rule of \cref{sec:bounds}, a
lexicographic test against the reversed sequence), propagation through the lag budgets, the bounds
$b_k$ on each $C_k^2$ that $\E\le T$ implies (\cref{thm:prop}, \cref{sec:branching}), and the
bound (\cref{thm:perlag,thm:pair}) at the current threshold $T$, then the exact energy if the node
is a leaf, and otherwise one sampler call on the node Hamiltonian, whose best sample is an
incumbent candidate. The children of the master are the nodes with $m_0$ boundary spins fixed on
each side, which represent the symmetry classes of
\citep{mertens1996exhaustive,packebusch2016labs}; we call them the class roots. We use
$m_0=\min(7,\lfloor n/2\rfloor)$, so that for $n\le14$ the class roots are leaves (even $n$) or
have a single free spin (odd $n$), and $m_0=7$ for every $n\ge14$. Every other node
branches on the spin chosen by the conflict values of its own samples.
After each node, $\UB$ and the certified lower bound are recorded with the time. The run ends
either when the lowest bound in the queue exceeds $T$, in which case $\UB$ is optimal, or at a
node or time limit, with a certified gap (\cref{cor:labs}).

\begin{algorithm}[t]
\caption{QCBB for LABS}
\label{alg:main}
\begin{algorithmic}[1]
\State $Q\gets\{(\LB(\text{master}),\text{master})\}$; $\UB\gets+\infty$; $T\gets+\infty$
\While{$Q\ne\emptyset$ and $\min_Q\LB\le T$}
  \State take $N$ with the lowest key from $Q$ (ties: the deeper node, then the older one)
  \State \textbf{if} $N$ fails the reversal rule (\cref{thm:sym}) \textbf{continue}
  \State force spins by the lag budgets until no spin changes (\cref{thm:prop}); \textbf{if} infeasible \textbf{continue}
  \State compute $R_k$, $\mu_k$ (\cref{thm:perlag}), $m_k$ and $\LB(N)$ (\cref{thm:pair}); \textbf{if} $\LB(N)>T$ \textbf{continue}
  \State \textbf{if} $U=\emptyset$: \textbf{if} $\E<\UB$: $\UB\gets\E$, $T\gets\UB-\Delta$; \textbf{continue}
  \State sample $\Hnode$ with fixed-angle QAOA; \textbf{if} the best sample has $\E<\UB$: $\UB\gets\E$, $T\gets\UB-\Delta$
  \State \textbf{if} $N$ is the master: children $\gets$ class roots
  \State \textbf{else}: $\chi\gets\nu P$ from the samples of $N$; $i^*\gets\arg\max\chi$ over the two outermost free spins; children $\gets$ $N$ with $s_{i^*}=\pm1$
  \State add each child $C$ with key $\max(\LB(N),\LB(C))\le T$ to $Q$
  \State record $(t,\UB,\LB_{\mathrm{cert}})$ if either bound changed
\EndWhile
\State \textbf{if} $Q$ has no key $\le T$: report $s_{\UB}$ as optimal; else report $s_{\UB}$, $\LB_{\mathrm{cert}}$ and the gap (\cref{thm:framework}(e),(f))
\end{algorithmic}
\end{algorithm}

\subsection{Quantum solution}
\label{sec:sampler}

The sampler is fixed-angle QAOA. It maps a node Hamiltonian $\Hnode$ on $F$ free spins, that
is, on $F$ qubits, to $q$ samples as in QCBB Sec.~3.1: distinct assignments
\begin{equation*}
  \hat s^\ell\in\{-1,+1\}^F\quad\text{with counts }\omega_\ell,\qquad \sum_\ell\omega_\ell=q .
\end{equation*}
Each sample is completed to a full sequence $s^\ell$, whose energy $\E(s^\ell)$ is computed
exactly in integers; the completion is an incumbent candidate, and it also gives the violation
matrix and the weights that select the branching spin (\cref{sec:branching}). The incumbent is
the best energy seen, $\UB=\E(s_{\UB})$, so $\UB$ is always attained. The sampler is called once
at every node that reaches it, from the master ($n$ qubits) and the class roots ($F=n-2m_0$
qubits) down to nodes with a single free spin, with the same number of shots $q$ at every node
(for the master at $n\ge27$ see \cref{sec:setup}). By \cref{thm:framework}, no choice made in
this section affects the validity of a proof.

\begin{figure}[tbp]
\centering
\includegraphics[width=\textwidth]{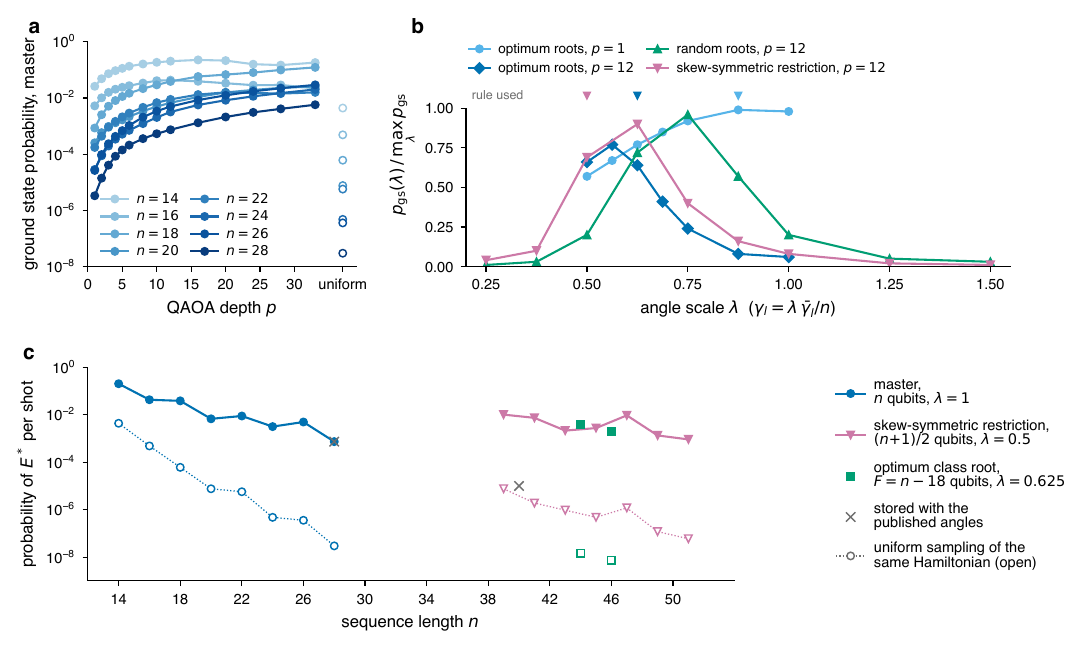}
\caption{The fixed-angle QAOA sampler.
(a) Ground state probability of the master Hamiltonian against the depth $p$ with the
transferred angles ($\lambda=1$) for $n=14$--$28$; open symbols in the right-hand column: uniform
sampling (ground states over $2^n$). (b) Angle scale: geometric mean over Hamiltonians of the
ground state probability at $\lambda$ relative to its maximum on the grid, for class roots whose
ground state is the global optimum ($22$ roots with $F=8$--$24$ free spins, that is, qubits) at
$p=1$ and $12$, random class roots ($24$, $F=12$--$24$) at $p=12$ and the skew-symmetric
restriction ($6$,
$n_{\mathrm s}=11$--$21$) at $p=12$; triangles mark the rule used ($\lambda=0.875$ at $p=1$ and
$0.625$ at $p=12$ for node Hamiltonians, $0.5$ for the skew-symmetric restriction). (c)
Probability of a sequence of energy $E^*$ per shot at $p=12$: the master ($n=14$--$28$; crosses
at $n=28,40$: values stored with the published angles, from the literature, the $n=28$ values
coincide), the skew-symmetric restriction on $n_{\mathrm s}=(n+1)/2$ qubits for the odd
$n=39$--$51$ where the skew-symmetric optimum is $E^*$, and the optimum class roots of $n=44,46$
($m_0=9$, $F=26,28$); open symbols: uniform sampling of the same Hamiltonian.}
\label{fig:qaoa}
\end{figure}

QAOA measures the depth-$p$ state $|\psi_p\rangle$ of the node Hamiltonian in the $Z$ basis. The state and the circuit are those of \citep{shaydulin2024qaoa}, in the angle
convention where each monomial $w\,Z_S$ of $\Hnode$ contributes the phase $e^{-i\gamma_l w Z_S/2}$
in layer $l$, and the stored fixed angles are those of the QOKit implementation
\citep{lykov2023qokit} (file \nolinkurl{qokit/assets/QAOA_with_fixed_parameters_p_opt.json}),
which originate from \citep{shaydulin2024qaoa}. There is no variational loop, because the fixed angles for the master
Hamiltonian transfer across lengths as $\gamma_l=\bar\gamma_l/n$, $\beta_l=\bar\beta_l$ with
$(\bar\gamma,\bar\beta)$ independent of $n$. A node Hamiltonian, however, has $1$- and $3$-spin
terms that the master does not have, and no transfer rule is known for it. We therefore rescale
the published angles, $\gamma_l=\lambda\,\bar\gamma_l/n$, $\beta_l=\bar\beta_l$, with $n$ the
sequence length (not $F$); for the master with $\lambda=1$ this is the published rule. For node
Hamiltonians we chose $\lambda$ on class roots with $m_0=6$, $8$ and $9$ boundary spins whose
ground state is the global optimum, since only these can
supply $\UB=E^*$; the runs use $m_0=7$. This uses knowledge of $E^*$, so $\lambda$ was then frozen for every run and
every instance; the proof is valid for any $\lambda$ (\cref{thm:framework}). On $22$ such roots
($F=8$--$24$ free spins, $m_0=6,8$), the geometric mean of the ground state probability relative
to the best $\lambda$ on a grid at $p=12$ peaks at $\lambda=0.5625$ with $0.77$, and is $0.66$
at $\lambda=0.5$, $0.64$ at the adopted $\lambda=0.625$, $0.24$ at $\lambda=0.75$ and $0.06$ at
$\lambda=1$, whereas random class roots prefer $\lambda\approx0.75$
(\cref{fig:qaoa}(b)). At $F=26$ and $F=28$ ($m_0=9$) the best grid value was $\lambda=0.625$, with
ground state probabilities $3.9\times10^{-3}$ and $2.0\times10^{-3}$ against $1.5\times10^{-8}$
and $7.5\times10^{-9}$ for uniform sampling, while at $p=1$ the optimum is flat around
$\lambda=0.875$--$1$. We therefore use $\lambda=0.875$ at $p=1$ and $\lambda=0.625$ for $p\ge2$
(measured at $p=1,4,12$), the best value at the two largest roots rather than at the peak of the
geometric mean. These constants are empirical (\cref{sec:discussion}). Since $\lambda$ enters
only the samples, it can change a run only through the time at which $\UB$ reaches $E^*$ and,
once the incumbent is optimal, through the choice of the branching spin (\cref{prop:samples}). For
the second effect we measured, with the optimum supplied at the root, that uniform samples give
the same tree size as QAOA at $p=12$ at $n=24$ ($4{,}643$ nodes), nearly the same at $n=27$
($12{,}424$ against $12{,}433$), $1$--$6\%$ larger trees at $n=25$, $26$ and $28$ and a $21\%$
smaller tree at $n=30$ (\cref{sec:samples}).

The sampled distribution reproduces the ground state probabilities stored with the angles in the
same file of the QOKit implementation \citep{lykov2023qokit} to all printed digits ($0.1735587$
at $n=12$, $p=12$; $0.0205713$ at $n=16$, $p=4$), and \cref{fig:qaoa}(a) shows the ground
state probability of the
master against $p$ for $n=14$--$28$. One call at $p=12$ with $q=1{,}000$ shots takes $0.08$,
$1.2$ and $5.0$\,s on the master at $n=20$, $24$, $26$, and its cost grows like $2^F$; but most
nodes of a proof have far fewer free spins than the master, so the mean call over a proof takes
$1.0$--$2.9$\,ms at $p=12$ and $0.7$--$1.5$\,ms at $p=1$ for $n=22$--$30$. In these runs the
cost of a call is nearly independent of the number of shots, because the state is computed once
and the shots are drawn from it (\cref{sec:setup}). On the master at $n=16$--$26$, one call at $p=12$
returns $E^*$ (seed~1), the ground state probability per shot being $3.2\times10^{-3}$ at $n=24$
(\cref{fig:qaoa}(a)), whereas at $p=1$ the master call misses $E^*$ at $n=18$, $19$ and
$22$--$26$. The gate count of the circuit, with each phase $e^{-i\gamma w Z_S/2}$ as a CNOT
ladder around one $R_Z$ rotation, is given in \cref{sec:hardware}.

For odd $n=2n_{\mathrm s}-1$, QAOA also samples the skew-symmetric restriction at the master
node. The skew-symmetric sequences (\cref{sec:intro}) are determined by their first
$n_{\mathrm s}$ spins and have $C_k=0$ for every odd $k$ \citep{golay1972class}, so their energy
is a quartic Ising Hamiltonian on $n_{\mathrm s}=(n+1)/2$ qubits with about $6.5$ times fewer
CNOTs per layer than the master ($536$ versus $3{,}890$ at $n=21$). The completed skew-symmetric
sequences are read as master samples; they are feasible, so the best one is a valid incumbent,
while the restriction enters no bound. The skew-symmetric optimum equals the optimum for $14$ of
the $24$ odd $n$ in $19$--$65$ \citep{packebusch2016labs} and differs for
$n=19,23,25,31,33,35,37,61,63,65$, but since the restricted samples are merged with samples of
the master itself, the incumbent is never worse than without the restriction. With QAOA at $p=12$
and $\lambda=0.5$ (the best grid value for the restriction), the probability of $E^*$ per shot is
$9\times10^{-4}$ to $10^{-2}$ at $n_{\mathrm s}=20$--$26$ ($n=39$--$51$, where the skew-symmetric
optimum is $E^*$; \cref{fig:qaoa}(c)), and at $n=27$ and $29$ the restriction on $14$ and $15$
qubits returns $E^*$ in the first call.

\subsection{Branching and propagation}
\label{sec:branching}

Branching splits a node into two children that fix one free spin to $+1$ and to $-1$, so the
children partition the completions of the node, which is condition (v); the master is the
exception, since its children are the class roots (\cref{sec:bounds}). QCBB chooses the spin
from conflict values computed from the samples of the node. No optimality claim depends on this
choice, because \cref{thm:framework} holds for any branching spin, and both children enter the
queue, so their order is set by their bounds and not by the samples.

QCBB computes the conflict values from the violations of the rows of $Ax=b$ by the samples,
but LABS has no linear rows. The constraints attached to a node are
instead the \emph{lag budgets}: in a completion with $\E\le T$, each $C_k^2$ can use at most the
part $b_k$ of $T$ that remains after the other lags have taken the least the bound allows them
(\cref{eq:budget}, \cref{app:prop}). Every completion with $\E\le T$ satisfies them. They are
node-specific and tighten as $\UB$ improves, and they are implicit in $\Hnode$, since a sample
that violates a budget already has $\Hnode>T$, so no penalty term is added. Let the samples of
the node, completed to full sequences, be the distinct sequences $s^\ell$ with counts
$\omega_\ell$, $q=\sum_\ell\omega_\ell$. Sample $\ell$ violates lag $k$ if $C_k(s^\ell)^2$
exceeds the budget $b_k$, and the violation matrix $V$, the weights $\nu$, the correspondence
matrix $P$ and the conflict values $\chi$ are
\begin{equation}
  \begin{gathered}
  V_{k\ell}=[\,C_k(s^\ell)^2>b_k\,],\qquad
  \nu_k=\frac1q\sum_\ell V_{k\ell}\,\omega_\ell,\\
  P_{ki}=[\,i-k\in\Phi\ \text{or}\ i+k\in\Phi\,],\qquad
  \chi_i=\sum_k\nu_kP_{ki},
  \end{gathered}
  \label{eq:conflict-main}
\end{equation}
for the lags $k=1,\dots,n-1$ and the free spins $i\in U$. So $\nu_k$ is the fraction of the
shots that violate the budget of lag $k$, $P$ marks a free spin for lag $k$ when it has a fixed
partner at that lag, so that it enters the linear part of $C_k$ (\cref{eq:sub}), and $\chi_i$
adds up the violation rates of the lags in which spin $i$ has a fixed partner. A node without
samples uses $\nu\equiv1$. These are QCBB eqs.~(5)--(8) with the lag budgets in place of the
rows, and $P$ is the analog of the nonzero pattern of the constraint matrix there; marking every
spin that appears in $C_k$ instead would make $\chi$ even flatter, because $C_k$ depends on every
spin with a lag-$k$ partner inside the sequence. The formal definition is \cref{def:conflict} in
\cref{app:conflict}.

The branching spin $i^*$ is the spin with the largest conflict value among a set of branching
candidates, the two outermost free spins: the leftmost and the rightmost one. Ties go to the free
spin nearer to an end, and between two spins at the same distance from the ends to the left one
if the number of fixed spins of the node is even and to the right one if it is odd. On LABS the
conflict values behave simply, and this motivates the candidate set. On a node whose fixed spins
form a prefix of $a$ spins and a suffix of $r$ spins, with a free block longer than both, the two
candidates have fixed partners at the same lags except for $|a-r|$ lags each, so the rule compares
the violation rates of the lags that differ: for $a=r+1$ the conflict value of the leftmost free
spin minus that of the rightmost one is $\nu_a-\nu_{n-1-r}$, a comparison of two lags (the lag
from the first spin to the leftmost free spin and the lag from the first spin to the rightmost
free spin), and for $a=r$ the two conflict values are equal for every $\nu$ and the tie rule
decides. With $\nu\equiv1$ the maximum
over all free spins is attained at an outermost free spin, so the branching order is that of
\citep{mertens1996exhaustive,packebusch2016labs}, which fixes the spins from the two ends inward,
and the restriction to the two outermost spins is the same rule evaluated on two spins instead of
$F$ (\cref{prop:frontier}). With the node's samples, $\nu$ decides which of the two outermost
free spins is fixed, so the samples change the search tree in two ways only: through the time at
which $\UB$ reaches the optimum, and through this choice at each branched node
(\cref{prop:samples}). Neither proposition is needed for optimality; both are proved in
\cref{app:conflict}, and \cref{sec:samples} measures the effect of the samples.

Substituting the fixed spins into each $C_k$ (QCBB, eqs.~(10)--(16)) leaves a constant, a linear
part over the free spins that have a fixed partner at lag $k$, and the products of two free spins
(\cref{eq:sub}, \cref{app:bounds}). The node Hamiltonian is the sum of the squares, so sample
energies are exact energies of full sequences, and every bound of \cref{sec:bounds} is a bound on
$\min_{\hat s}\Hnode(\hat s)$. Both the bound and the propagation rule use the \emph{range}
$R_k$ of $C_k$, the set of values that $C_k$ takes over all completions of the node, which is an
arithmetic progression of step $2$ or $4$, computed exactly by one scan of the lag-$k$ products in $O(n-k)$
integer operations (\cref{thm:perlag}, \cref{fig:chains}). Its smallest modulus $\mu_k$ gives
$C_k^2\ge\mu_k^2$, so $\LB_0=\sum_k\mu_k^2$ is the per-lag bound, which does not decrease from a
node to its children and lies on the energy lattice. \Cref{thm:perlag} is the ``tight bound'' of
\citep[Sec.~3.1.5]{packebusch2016labs} for the contiguous free block left by the shells of that
algorithm (the pairs of spins fixed from the two ends), whereas our proof covers arbitrary fixed
sets, which propagation produces (\cref{app:bounds}).

QCBB propagates forced variables through the constraints and detects contradictions (QCBB
Sec.~3.2.4); in LABS the constraints are the lag budgets. The budget of lag $k$ leaves the set
$A_k$ of values in $R_k$ that $C_k$ can still take in a completion with $\E\le T$. If some $A_k$
is empty, the node has no such completion. If $A_k$ is the single smallest (or largest) value of
$R_k$, every free chain of lag $k$ (a maximal run of free lag-$k$ products, \cref{app:bounds})
must attain its own minimum (or maximum); for a free chain with a fixed spin at one end this
forces all of its products to one sign, so that its free spins are determined uniquely from that
end and are fixed (\cref{app:prop}). The forcings are repeated until no spin changes, which takes
at most $|U|$ rounds, and the ranges are recomputed after the
first lag that forces anything, so that every forcing uses the sets of the current assignment. A
spin forced to both values, or $\LB>T$, shows that the node has no completion with $\E\le T$.
\Cref{thm:prop} (\cref{app:prop}) states this rule exactly and proves that it removes only
completions with $\E>T$, which is condition (ii), and the remark after it records why the exact
test for a single value matters: a rule written on the plain range forces spins wrongly.

\subsection{Bounding, pruning and stopping}
\label{sec:bounds}

QCBB selects the open node with the lowest bound (QCBB Sec.~3.3), and so do we, with ties going
to the deeper node and then to the node that entered the queue first. A child enters the queue
with the key $\max(\LB(N),\LB(C))$ of its parent $N$ and itself, and only if this key is at most
$T$; here $\LB(C)$ is the bound of \cref{thm:pair} on the child as created, before propagation,
and it is recomputed after propagation when the child is evaluated (\cref{alg:main}, lines 6 and
11). Any order is valid for \cref{thm:framework}.

The Goemans--Williamson bound of QCBB (eqs.~(17)--(19)) is far too loose for LABS
(\cref{sec:intro}), so we use instead an integer bound built from the ranges $R_k$. The per-lag
bound $\LB_0$ of \cref{sec:branching} treats each lag on its own; the pair congruence
$C_k+C_{n-k}\equiv n\pmod 4$ of \citep{wiggenbrock2010labs,packebusch2016labs} couples the lags
$k$ and $n-k$, and
the bound takes for each pair the smallest $C_k^2+C_{n-k}^2$ over the values in the two ranges
whose sum has the right residue. This costs $O(1)$ per pair, because each residue class of
an arithmetic progression of step $2$ or $4$ is again a progression (\cref{eq:lb},
\cref{thm:pair}, \cref{app:bounds}). Every completion of the node satisfies $\E\ge\LB\ge\LB_0$,
so $\LB$ satisfies condition (i); at the root, for $n\equiv2\pmod 4$, $\LB=3n/2-2$, which equals
the optimum for $n=6,10,14,18$, while for $n=22,26,30$ it is $31,37,43$ against $E^*=39,45,59$, a
root gap of $8,8,16$ on the lattice of step $8$.

QCBB prunes a node when its lower bound is above the incumbent (QCBB Sec.~3.5); with the energy
lattice, a node is pruned when $\LB>T=\UB-\Delta$ (\cref{lem:lattice}(c)). A node is infeasible
for the search at threshold $T$ when it has no completion with $\E\le T$, which the propagation
rule of \cref{sec:branching} reports when a set $A_k$ is empty, when two forced values contradict,
or when $\LB>T$. This takes the role of QCBB's infeasibility test (QCBB App.~C), without $M$,
and together with the bound test it gives condition (ii).

QCBB has no symmetry rule, but LABS has a group of order~8 that leaves the energy unchanged,
generated by negation $s_i\mapsto -s_i$, alternation $s_i\mapsto(-1)^{i-1}s_i$ and reversal
$s_i\mapsto s_{n+1-i}$ \citep{packebusch2016labs}, and a symmetry rule may discard a node if no
completion lies in a set $K$ that meets every orbit. Negation and alternation together bring every
sequence $s$ to one element $\kappa(s)$ with $s_1=s_2=+1$, and the orbit of $s$ meets this set in
the set $\{\kappa(s),\kappa(\rho s)\}$, where $\rho s$ is the reversal of $s$ (the two elements
coincide for some orbits), so fixing $s_1=s_2=+1$ satisfies condition (iii) and
gives $4\times$ fewer leaves. The \emph{reversal rule}, a lexicographic leader test as in
\citep[Sec.~2.6]{prestwich2013labs}, keeps, of these two elements, the lexicographically smaller one: on a node with $s_1=s_2=+1$ and a prefix and a suffix of $m\ge2$
fixed spins each, the first $m$ entries of the reversed element depend only on the suffix, and if
the comparison with the prefix is decided within these entries and the reversed element is
smaller, the node is discarded. \Cref{thm:sym} (\cref{app:prop}) proves both rules. The rule is
evaluated with $m=m_0$ at every node whose two end blocks are fixed, but every descendant of a
class root has the same $m_0$ boundary spins on each side, so the test repeats the verdict of
the root and acts once, on the class roots, the children of the master; a test with a longer
prefix and suffix at deeper nodes would be valid by \cref{thm:sym}(b) but is not applied.
Roots that fail the rule are dropped, which keeps $2{,}112$ of the $4{,}096$ roots at $m_0=7$ for
every $n$. For odd $n$ this is the class count $c(7)$ of \citep[Sec.~3.2,
eq.~(26)]{packebusch2016labs}; for even $n$ that count is $2{,}080$, so $32$ of the kept roots
represent a class that another kept root also represents, which is harmless for correctness by
condition (iii) and costs at most $32$ extra nodes. The class count formula is not used; the
roots are enumerated and filtered by the rule.

The run stops when the lowest key in the queue exceeds $T=\UB-\Delta$, because then no completion
of an open node can improve $\UB$, and the incumbent is optimal (\cref{thm:framework}(c), QCBB
Sec.~3.6.1); a leaf or a sample with $\E\le T$ instead lowers $\UB$ and $T$, and the run
continues. At any time, the certified lower bound is the smaller of $\UB$ and the lowest key in
the queue, rounded up to the energy lattice (\cref{thm:framework}(e),(f); QCBB Sec.~3.6.2), so a
run with a node or time limit stops early and reports the incumbent with this bound and the gap.

\subsection{Correctness}
\label{sec:correctness}

The guarantee is one theorem about the framework, stated and proved in \cref{app:framework}. It
refers to a lower bound $\LB(N)$ on $f$ over the completions of a node $N$, to a propagation rule
that maps a node with threshold $T$ either to ``infeasible'' or to a node with more fixed
spins, and to a symmetry rule for a group under which $f$ is invariant. Under conditions
(i)--(v), the following holds (\cref{thm:framework}). The search tree is finite
and each node is evaluated once, and if some $x$ has $f(x)\le T$, the search at threshold $T$
reaches a leaf with $f\le T$; hence a search at $T=\UB-\Delta$ that ends with no leaf of value
$\le T$ proves $\UB=\min f$, and a completed search at any lattice threshold $T'$ with no leaf of
value $\le T'$ proves $\min f\ge T'+\Delta$.
During a search, the smallest bound among the open nodes, capped by $T+\Delta$, is a valid lower
bound on $\min f$, and the larger of this open-node bound and the highest completed threshold
plus $\Delta$ is the certified lower bound $\LB_{\mathrm{cert}}\le\min f\le\UB$, so that
$(\UB-\LB_{\mathrm{cert}})/\UB$ is a valid gap. Sample data enter only as incumbent candidates,
re-evaluated exactly, and as the branching variable, so no part of the theorem depends on them.

QCBB's original instance, the binary linear program with the certified MaxCut bound, variable
bound propagation and the infeasibility test, satisfies the conditions with the trivial group and
$\Delta=1$; we make no new claim about it (\cref{app:framework}). For LABS, the bound of
\cref{sec:bounds} gives (i), the propagation rule of \cref{sec:branching} gives (ii), the
restriction $s_1=s_2=+1$ with the reversal rule gives (iii), the energy lattice of \cref{sec:labs}
gives (iv) and binary branching on any free spin gives (v). So \cref{alg:main} returns a proven
optimum or a valid gap, whatever the sampler returns (\cref{cor:labs}).

\section{Results}
\label{sec:results}

\subsection{Setup and metrics}
\label{sec:setup}

We use the LABS instances of QOBLIB \citep{koch2026qoblib}, labs002--labs100 for $n=2,\dots,100$,
and compare with the optimal values listed there. All runs use the algorithm of \cref{sec:qcbb}
with $m_0=7$ ($2{,}112$ class roots after the reversal rule, on $F=n-14$ free spins), the two
outermost free spins as branching candidates, $q=1{,}000$ shots per node, seed~1 and a limit of
one hour per run, which no run reached; \cref{app:seeds} repeats the QAOA runs at $p=12$ with
seeds $1$--$5$ and no limit. A run is one process on an Apple M4 Pro (12 cores, 24\,GB memory,
macOS 27.0.1) that uses all of its cores. There are three series. Two use QAOA, at depth $p=12$
and at $p=1$, with the fixed angles of \citep{shaydulin2024qaoa} and the angle rescaling of
\cref{sec:sampler} ($\lambda=1$ for the master, $0.625$ at $p=12$ and $0.875$ at $p=1$ for the
other nodes, $0.5$ for the skew-symmetric restriction), for $n=2$--$30$. The third series is the
classical reference: the same algorithm with simulated annealing (SA)
\citep{kirkpatrick1983annealing} in place of QAOA, on $\Hnode$ with single-spin flips of the free spins, where each shot is one independent
Metropolis annealing run of $1{,}000$ sweeps with a geometric inverse-temperature schedule from
$\beta=0.02$ to $\beta=2.0$ that returns the best state it visited; the energy change of a flip
is computed in $O(n)$ from the cached autocorrelations $C_k$, so the cost is linear
in the number of shots, and the shots of a call run in parallel. A call of $1{,}000$ shots takes
$47$--$86$\,ms on the master at $n=16$--$26$ and $14$--$26$\,ms on average over the nodes of a
proof at $n=19$--$26$; SA was run for $n=2$--$26$, and \cref{sec:samples} adds runs with
smaller budgets per call. Its parameters were fixed once and not
tuned, and memetic tabu search \citep{gallardo2009memetic,boskovic2017labs} is the stronger
classical alternative that would lower the SA numbers further; we use SA because it is the
simplest sampler of the same kind as QAOA, one assignment per shot from a distribution
concentrated on low energies. On the master at $n=16$--$26$, one call of SA returns $E^*$
(seed~1), as does one call of QAOA at $p=12$ (\cref{sec:sampler}). All numbers in this section
are measured on this machine unless they are labeled (estimate).

The QAOA circuits were run on an exact noiseless statevector simulator, which computes the state
of a circuit once and draws the shots from it; we set its limit to $26$ qubits, below what the
memory of the machine would allow. At $n=27$--$30$ the master, the full problem on $n$ spins, exceeds
this setting; since the master is split into its symmetry classes, it needs no samples to branch,
so there it is split without being sampled (marked in \cref{tab:backends} and
\cref{fig:backends}), and for odd $n$ its skew-symmetric restriction on $(n+1)/2$ qubits is
sampled and supplies incumbent candidates. Every other node has at most $n-14\le16$ free spins
and is sampled. Where a noise model is used, \cref{sec:hardware} names it.

The following consistency checks were made. The ranges $R_k$ equal the sets of $C_k$ values
enumerated over all completions at $600$ random nodes with $n\le13$ and arbitrary fixed sets. At
$1{,}500$ random nodes, every completion with $\E\le T$ satisfies the lag budgets and survives
propagation. The algorithm returns the optimum for $n\le14$ also with a sampler that returns the
all-$(+1)$ assignment, as \cref{thm:framework} requires.

The primary cost measure is the number of sampler calls of a run and, with it, the number of
shots, $1{,}000$ per call; neither depends on how the sampler is executed. The runtime is the
wall-clock time from the start of a run to the proof of optimality (or to the stop), and it is a
time of the simulation, as is the primal-dual integral defined below. It is split into the sampler time, spent in sampler calls (QAOA
or annealing, and the exact energies of the samples), and the time of the node loop
itself (queue, bounds, propagation, branching). The second metric is the primal-dual integral. Let $\UB(t)$ be the
incumbent energy at time $t$ after the start of the run and $\LB(t)$ the certified lower bound at
that time (\cref{thm:framework}(e),(f)), both of which the run records after every node, with the
time, whenever one of them changes. We use the primal gap function and primal integral of
\citep{berthold2013primal} with one change: the certified lower bound $\LB(t)$ takes the place of
the optimum, as in QCBB Sec.~4.3 and in the dual gap of \citep{berthold2018phases}. The
primal-dual gap is thus
\begin{equation}
  \begin{gathered}
  g(t)=\begin{cases}1 & \text{if }\UB(t)=+\infty\text{ or nothing is recorded yet},\\
  \dfrac{\UB(t)-\LB(t)}{\UB(t)} & \text{otherwise},\end{cases}\\
  \mathrm{PDI}=\int_0^{t_{\mathrm{end}}} g(t)\,dt,
  \end{gathered}
  \label{eq:pdi}
\end{equation}
where $t_{\mathrm{end}}$ is the end of the run; the bounds are recorded after each node, so the
first record follows the first node (the master and its sampler call), and recording the root
bound at $t=0$ would change nothing, because the gap is $1$ without an incumbent. Being the area
under the gap curve, the integral rewards a good incumbent and a strong bound that appear early,
and a run that proves optimality has $g=0$ at its end. For LABS every energy is at least
$C_{n-1}^2=1$, so the $0/0$ case of \citep{berthold2013primal} cannot occur. Because the integral
is measured in seconds, a slower sampler increases it even if its samples are as good; the
time-normalized integral $\mathrm{PDI}/t_{\mathrm{end}}$, the mean gap over the run, is reported
with it. Both metrics are measured; neither enters a bound or a proof.

\subsection{Proven optima and convergence}
\label{sec:experiments}

Every run proved the QOBLIB optimum: labs002--labs030 with QAOA at $p=12$ and at $p=1$, and
labs002--labs026 with SA. \Cref{tab:backends} gives, for each run at $n=16$--$30$, the node
count, the number of sampler calls (with $1{,}000$ shots each), the runtime and the PDI, and in
its lower block the two reference series of \cref{sec:samples}; for $n\le15$ every series
needs one sampler call, at most $6$ nodes and at most $0.07$\,s. A node is counted when it is taken from the queue and evaluated, whether
it is then dropped by the reversal rule, by propagation or by the bound, sampled and branched, or
a leaf; children whose key exceeds $T$ never enter the queue and are not counted. A sampler call
is counted when the sampler returns samples for a node. The smallest proof in the table is at
$n=18$, where the master samples of
SA and of QAOA at $p=12$ reach $E^*$ and every class root has $\LB>T$ (at the root,
$\LB=3n/2-2=E^*$ for $n=18$, \cref{thm:pair}), so the proof is a single node; the largest, $n=29$
at $p=12$, evaluated $443{,}140$ nodes with $256{,}140$ sampler calls ($2.6\times10^8$ shots)
in $415$\,s. At
$n=23$--$30$, $48$--$61\%$ of the nodes make a sampler call, while the others are dropped by the
reversal rule, propagation or the bound before the call. The mean gap $\mathrm{PDI}/t_{\mathrm{end}}$
of \cref{eq:pdi} is close to $1$ for $n\le18$, where a run is little more than the master call,
and at $n=22$--$30$ it is $0.30$--$0.58$ for QAOA at $p=12$, $0.30$--$0.39$ at $p=1$ and
$0.18$--$0.35$ for SA ($n=22$--$26$).

\begin{table}[tbp]
\centering\footnotesize\setlength{\tabcolsep}{2.9pt}
\begin{adjustbox}{max width=\textwidth}
\begin{tabular}{rr rrrr rrrr rrrr}
\toprule
 & & \multicolumn{4}{c}{QAOA, $p=12$} & \multicolumn{4}{c}{QAOA, $p=1$} &
 \multicolumn{4}{c}{SA}\\
\cmidrule(lr){3-6}\cmidrule(lr){7-10}\cmidrule(lr){11-14}
$n$ & $E^*$ & nodes & calls & runtime & PDI & nodes & calls & runtime & PDI & nodes & calls & runtime & PDI\\
 & & & & (s) & (s) & & & (s) & (s) & & & (s) & (s)\\
\midrule
16 & 24 & 61 & 1 & 0.06 & 0.06 & 61 & 1 & 0.04 & 0.04 & 61 & 1 & 0.07 & 0.07\\
17 & 32 & 319 & 13 & 0.07 & 0.06 & 319 & 13 & 0.05 & 0.04 & 319 & 13 & 0.15 & 0.11\\
18 & 25 & 1 & 1 & 0.04 & 0.04 & 53 & 1 & 0.04 & 0.03 & 1 & 1 & 0.05 & 0.05\\
19 & 29 & 411 & 73 & 0.14 & 0.10 & 443 & 99 & 0.11 & 0.07 & 411 & 73 & 1.04 & 0.38\\
20 & 26 & 328 & 130 & 0.22 & 0.15 & 328 & 130 & 0.15 & 0.09 & 328 & 130 & 2.25 & 0.94\\
21 & 26 & 381 & 129 & 0.29 & 0.21 & 381 & 129 & 0.18 & 0.10 & 381 & 129 & 2.32 & 0.79\\
22 & 39 & 1,777 & 730 & 1.05 & 0.46 & 1,968 & 915 & 1.00 & 0.31 & 1,777 & 730 & 13.71 & 2.89\\
23 & 47 & 14,604 & 7,519 & 8.42 & 3.14 & 14,736 & 7,611 & 8.09 & 2.76 & 14,604 & 7,519 & 148.21 & 47.88\\
24 & 36 & 4,643 & 2,598 & 4.49 & 2.39 & 13,348 & 7,575 & 9.24 & 3.56 & 4,388 & 2,425 & 55.82 & 19.73\\
25 & 36 & 6,144 & 3,259 & 6.84 & 3.64 & 14,352 & 9,400 & 12.57 & 4.81 & 5,522 & 2,853 & 70.20 & 19.61\\
26 & 45 & 7,084 & 3,392 & 10.39 & 6.00 & 9,341 & 5,572 & 8.77 & 3.15 & 6,631 & 3,166 & 82.89 & 14.83\\
$27^\dagger$ & 37 & 12,433 & 7,025 & 12.17 & 3.68 & 12,403 & 6,998 & 11.60 & 3.53 & -- & -- & -- & --\\
$28^\dagger$ & 50 & 134,039 & 76,236 & 120.58 & 43.82 & 138,186 & 78,662 & 118.26 & 43.41 & -- & -- & -- & --\\
$29^\dagger$ & 62 & 443,140 & 256,140 & 414.75 & 146.69 & 459,708 & 265,898 & 401.25 & 142.37 & -- & -- & -- & --\\
$30^\dagger$ & 59 & 182,737 & 110,867 & 214.48 & 65.56 & 186,980 & 113,251 & 188.34 & 59.69 & -- & -- & -- & --\\
\bottomrule
\end{tabular}
\end{adjustbox}

\medskip
\begin{adjustbox}{max width=\textwidth}
\begin{tabular}{rr rrrr rrrr}
\toprule
 & & \multicolumn{4}{c}{uniform samples} &
 \multicolumn{4}{c}{optimum at root, no samples}\\
\cmidrule(lr){3-6}\cmidrule(lr){7-10}
$n$ & $E^*$ & nodes & calls & runtime & PDI & nodes & calls & runtime & PDI\\
 & & & & (s) & (s) & & & (s) & (s)\\
\midrule
16 & 24 & 61 & 1 & 0.03 & 0.03 & 61 & 1 & 0.03 & 0.03\\
17 & 32 & 331 & 27 & 0.04 & 0.04 & 319 & 1 & 0.03 & 0.03\\
18 & 25 & 68 & 12 & 0.04 & 0.03 & 1 & 1 & 0.00 & 0.00\\
19 & 29 & 486 & 149 & 0.12 & 0.07 & 411 & 1 & 0.04 & 0.03\\
20 & 26 & 501 & 363 & 0.26 & 0.17 & 328 & 1 & 0.04 & 0.03\\
21 & 26 & 518 & 272 & 0.24 & 0.14 & 381 & 1 & 0.04 & 0.03\\
22 & 39 & 1,994 & 947 & 0.90 & 0.26 & 1,777 & 1 & 0.11 & 0.04\\
23 & 47 & 14,736 & 7,611 & 7.61 & 2.52 & 14,604 & 1 & 0.86 & 0.29\\
24 & 36 & 13,348 & 7,575 & 9.04 & 3.31 & 4,338 & 1 & 0.30 & 0.12\\
25 & 36 & 14,425 & 9,448 & 12.60 & 4.53 & 5,388 & 1 & 0.35 & 0.11\\
26 & 45 & 9,470 & 5,694 & 8.29 & 2.32 & 6,563 & 1 & 0.39 & 0.09\\
27 & 37 & 15,424 & 9,581 & 15.99 & 6.61 & 8,378 & 1 & 0.56 & 0.17\\
28 & 50 & 144,044 & 82,229 & 134.03 & 49.17 & 81,050 & 1 & 5.20 & 1.53\\
29 & 62 & 481,176 & 278,582 & 339.49 & 121.50 & 256,805 & 1 & 17.28 & 4.86\\
30 & 59 & 193,161 & 116,916 & 211.11 & 65.92 & 91,534 & 1 & 5.71 & 1.16\\
\bottomrule
\end{tabular}
\end{adjustbox}
\caption{Proofs of the QOBLIB optima with QAOA at $p=12$, QAOA at $p=1$ and SA as the sampler of
the same algorithm (settings of \cref{sec:setup}, $1{,}000$ shots per node, one process,
seed~1). Every run proves the QOBLIB optimum $E^*$. Nodes and sampler calls are counted as
defined in \cref{sec:experiments}. The runtime is the time to the proof; the
sampler time is $89$--$92\%$ of it for QAOA at $n=28$--$30$ and $97$--$99\%$ for SA at
$n=19$--$26$. $^\dagger$: master not sampled (\cref{sec:setup}); for odd $n$ its skew-symmetric
restriction is sampled instead, which counts as one call. --: SA was not run above $n=26$.
For $n\le15$ every run takes at most $6$ nodes and $0.07$\,s; the full range is in
\cref{fig:backends}(b). Lower block: the two reference series of \cref{sec:samples} with the
same settings and seed, run on the same machine in a later session: uniform random samples,
$1{,}000$ per node, and the known optimum supplied at the root with no samples afterwards. In
the latter the sampler is called below the root but returns nothing, so the calls column counts
the calls that returned a sample ($1$) and the runtime is that of the node loop with these empty
calls.}
\label{tab:backends}
\end{table}

\begin{figure}[tbp]
\centering
\includegraphics[width=\textwidth]{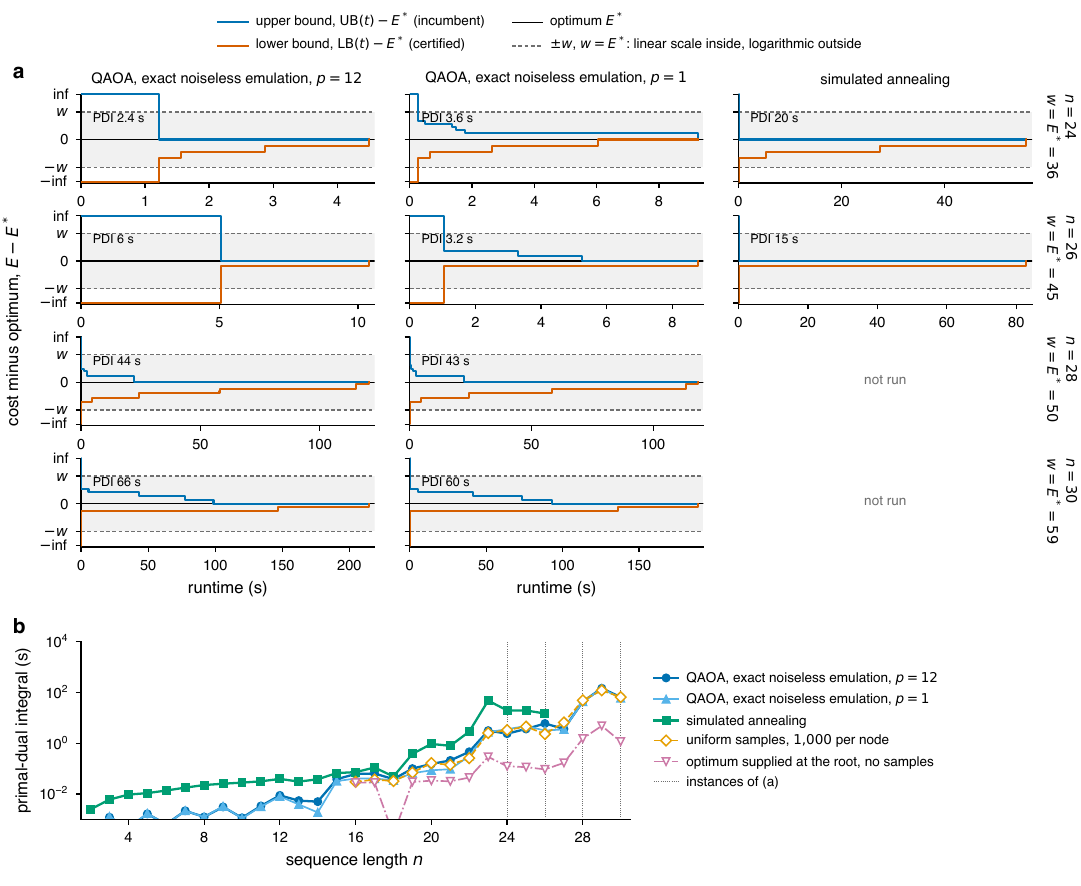}
\caption{Convergence of the bounds with QAOA at $p=12$, QAOA at $p=1$ and SA as the sampler of
the same algorithm (settings of \cref{sec:setup}, $1{,}000$ shots per node, one process, seed~1;
measured). (a) The blue line is the upper bound, the energy of the incumbent, and the vermilion
line the certified lower bound, both as functions of the runtime in seconds, on the instances
$n=24,26,28,30$ (rows) for the samplers (columns) QAOA with the fixed angles at $p=12$ and at
$p=1$, and simulated annealing (not run at $n=28,30$). A solid black line marks the optimum
$E^*$, aligned to $0$;
between the dashed lines at $\pm w$ the $y$ axis is linear, outside them logarithmic. LABS has no
infeasible sequences, so in place of QCBB's cost of the worst feasible solution we take the
half-width $w=E^*$ of the linear band, which then holds every bound between the trivial lower
bound $0$ and $2E^*$. The ticks inf and $-$inf mark the time before the first
incumbent and before the first certified bound ($g=1$ there). The PDI of the run, the area between
the curves with the gap normalized by $\UB(t)$ (\cref{sec:setup}), is written in each panel. At
$n=24,26$ the first sampler call, on the master, takes $1.2$ and $5.0$\,s at $p=12$. At
$n=28,30$ the master is not sampled (\cref{sec:setup}), and the incumbent descends as the nodes
below it are sampled. (b) PDI against $n$ for the three series
($n=2$--$30$ for QAOA, $2$--$26$ for SA) and, with broken lines and open symbols, for the two
reference series of \cref{sec:samples} at $n=16$--$30$ (uniform samples, and the optimum
supplied at the root with no samples afterwards, whose PDI at $n=18$, a single node, is below the
axis range); vertical lines mark the instances of (a). For
$n\ge27$ the master is not sampled (\cref{sec:setup}), so the first incumbent comes from the
skew-symmetric restriction (odd $n$) or from the first class root (even $n$). Every run
proves the QOBLIB optimum.}
\label{fig:backends}
\end{figure}

The two bounds converge as in a classical branch and bound (\cref{fig:backends}(a)). Where the
master is sampled ($n\le26$) and its samples reach $E^*$, the upper bound is optimal after the
first call, at $n=24$ after $1.2$\,s with QAOA at $p=12$ and after $0.1$\,s with SA, and the
rest of the run raises the certified lower bound, in steps, as the lowest key in the queue rises.
For $n\equiv2\pmod4$ the step $\Delta=8$ makes these steps coarse: at $n=26$ the lattice points
next to $E^*=45$ are $37$ and $53$, so the certified bound stays at $37$ until the last node with
key $37$ is closed, and then jumps to $45$. Where the master is not sampled, at $n=28$ and $30$
with QAOA, the first class root supplies the first incumbent within $0.06$\,s, and the incumbent
then descends as the nodes are sampled: at $n=28$ and $p=12$ through $74$, $70$, $62$ to $E^*=50$
at $22$\,s, and at $n=30$ through $91$, $83$, $75$, $67$ to $E^*=59$ at $99$\,s ($93$\,s at
$p=1$), while the proofs end at $121$ and $214$\,s. The PDI is well below the runtime because the
gap, normalized by $\UB(t)$, is far below $1$ for most of the run.

The number of sampler calls sets the cost. A proof at $n=22$--$26$ makes $730$--$7{,}519$ calls,
that is, $0.73$--$7.5$ million shots, with QAOA at $p=12$ and with SA, and $915$--$9{,}400$ calls
at $p=1$; a proof at $n=28$--$30$ makes $76{,}236$--$265{,}898$ calls, up to $2.7\times10^8$
shots, at $p=12$ and $p=1$. One sampler call per node makes the sampler time $89$--$92\%$ of the
runtime with QAOA at $n=28$--$30$, and $97$--$99\%$ with SA, whereas the node loop outside the
sampler costs $60$--$110\,\mu$s per node. In a profiled repeat of the $n=28$ run at $p=12$
(seed~1; the profiler about doubles the loop time), the node loop outside the sampler took
$169\,\mu$s per node, of which propagation took $1.8\%$ and the conflict values of the samples
$68\%$. The QAOA calls take $1.0$--$2.9$\,ms on average
at $p=12$ and $0.7$--$1.5$\,ms at $p=1$ ($n=22$--$30$), and the SA calls $14$--$26$\,ms with
$1{,}000$ shots ($n=19$--$26$), so at $n=26$ the proof takes $10.4$\,s at $p=12$, $8.8$\,s at
$p=1$ and $82.9$\,s with SA, with PDIs of $6.0$, $3.2$ and $14.8$\,s. At $n=28$--$30$ the runs at
$p=12$ and $p=1$ are within $14\%$ of each other in runtime and PDI, because neither samples the
master and the shorter $p=1$ calls are offset by $2$--$4\%$ more nodes. A least-squares fit of
the logarithm of the node count on $n$ over $n=20$--$30$ at $p=12$ gives a growth of $1.94$ per
unit of $n$ (one standard error of the slope: $1.78$--$2.12$; $95\%$ interval $1.59$--$2.37$;
residual scatter $\times/2.5$). The interval contains both the $1.73$
measured for the classical branch and bound of \citep{packebusch2016labs} and the $2$ of plain
enumeration, so eleven instances with this scatter do not determine the growth rate, and the
$1.73$ is in any case a runtime exponent of a different algorithm, measured with a different node
definition and over a different range of $n$. Gurobi,
on a binary quadratically constrained quadratic formulation, proved optimality up to $n=41$, in about 13 hours for that instance
\citep{koch2026qoblib}.

\subsection{What the samples contribute}
\label{sec:samples}

By \cref{prop:samples}, the samples change the search tree in two ways: through the time at which
$\UB$ reaches $E^*$, and through the choice between the two outermost free spins at each branched
node. To measure both, the same algorithm was run with two reference samplers in place of QAOA,
with the settings of \cref{sec:setup} and seed~1 (lower block of \cref{tab:backends},
\cref{fig:backends}(b)): uniform random samples, $1{,}000$ per node, and the known optimum
supplied at the root with no samples afterwards, so that $\nu\equiv1$ at every node below the
root, which is the fixed-threshold scheme of \citep{packebusch2016labs}, where the search starts
from the best energy found by a heuristic. In the second the sampler is called below the root but
returns nothing, so no shots are spent.

The first way accounts for the largest differences. At $p=1$ the master
call misses $E^*$ at $n=18$, $19$ and $22$--$26$, and the proof then takes up to $2.9\times$ more
nodes than at $p=12$, whose master call returns $E^*$ ($n=24$: $13{,}348$ against $4{,}643$). At
$n=24$, for instance, the incumbent of the $p=1$ run descends through $60$, $56$, $52$, $48$,
$44$ in the first $1.8$\,s and reaches $E^*=36$ only at the end of the run, at $9.2$\,s, so that
every node evaluated before then is judged at a larger threshold. Uniform samples behave like
QAOA at $p=1$: their trees coincide with the $p=1$ trees at $n=16$, $23$ and $24$ ($61$,
$14{,}736$ and $13{,}348$ nodes), are within $1.4\%$ of them at $n=22$, $25$ and $26$, up to
$53\%$ larger at $n=17$--$21$ ($501$ against $328$ at $n=20$), and $3$--$24\%$
larger at $n=27$--$30$; they reach $E^*$ after $24.6$\,s at $n=28$ and after $101$\,s at
$n=30$, against $22$ and $99$\,s with QAOA at $p=12$, and after $11.0$\,s at $n=29$, where QAOA
has it from the skew-symmetric restriction in the first call.

The second way is isolated by supplying the optimum at the root. With $\nu\equiv1$ below it the
tree is the smallest of all series at every $n\ge24$: $4{,}338$ nodes at $n=24$ against $4{,}643$
with QAOA at $p=12$ and $4{,}388$ with SA, and $81{,}050$ and $91{,}534$ nodes at $n=28$ and $30$
against $134{,}039$ and $182{,}737$ with QAOA at $p=12$, which reached $E^*$ only after $22$ and
$99$\,s; over $n=24$--$30$ the trees of QAOA at $p=12$ are $1.07$--$2.00$ times this sample-free
tree. To separate the branching choice from the time to $E^*$, we also ran uniform samples below
the optimum supplied at the root. This gives the same tree as QAOA at $p=12$ at $n=24$ ($4{,}643$
nodes), nearly the same at $n=27$ ($12{,}424$ against $12{,}433$), $1$--$6\%$ larger trees at
$n=25$, $26$ and $28$, and a $21\%$ smaller tree at $n=30$ ($143{,}708$ against $182{,}737$),
while against $\nu\equiv1$ with the same optimum at the root these sample-driven trees are
$7$--$15\%$ larger at $n=24$--$26$ and $48$--$76\%$ larger at $n=27$, $28$ and $30$. In the proven
regime the samples therefore shorten a proof only through $\UB$, and once the incumbent is optimal
the branching choice driven by the samples, whether from QAOA or uniform, enlarges the tree
against the rule $\nu\equiv1$, which fixes the spins from the two ends inward
(\cref{prop:frontier}). SA and QAOA at $p=12$, whose master calls both return $E^*$, visit
identical trees at $n=16$--$23$ and differ by $5$--$11\%$ at $n=24$--$26$ ($4{,}388$ against
$4{,}643$; $5{,}522$ against $6{,}144$; $6{,}631$ against $7{,}084$), because their samples weight
the lags differently and so fix different outermost spins at some nodes.

The SA series of \cref{tab:backends} has one budget per call, so the same algorithm was also run
with SA at $10$ shots of $100$ sweeps and at $100$ shots of $100$ sweeps per call
(\cref{tab:sabudget}). Every run proves the optimum. With $100\times100$ the first call returns
$E^*$ at every $n=16$--$26$, the trees are those of the $1{,}000\times1{,}000$ series at
$n\le23$ and within $3.1\%$ of them at $n=24$--$26$, with the same number of calls to within
$3.5\%$, and the runtime at $n=26$ is $2.1$\,s instead of $82.9$\,s. With $10\times100$ the first
call misses $E^*$ at $n=20$--$22$, $24$ and $25$, and the trees and the number of calls then grow
as with QAOA at $p=1$: $12{,}782$ and $12{,}788$ nodes at $n=24$ and $25$, where $E^*$ is reached
only at the end of the run, against $4{,}388$ and $5{,}522$; at $n=16$--$19$, $23$ and $26$ the
first call returns $E^*$ and the trees stay within $1.1\%$ of the $1{,}000\times1{,}000$ series.

\begin{table}[tbp]
\centering\footnotesize\setlength{\tabcolsep}{1.9pt}
\begin{adjustbox}{max width=\textwidth}
\begin{tabular}{rr rrrrr rrrrr rrr}
\toprule
 & & \multicolumn{5}{c}{SA, $10$ shots $\times$ $100$ sweeps} & \multicolumn{5}{c}{SA, $100$ shots $\times$ $100$ sweeps} &
 \multicolumn{3}{c}{SA, $1{,}000\times1{,}000$}\\
\cmidrule(lr){3-7}\cmidrule(lr){8-12}\cmidrule(lr){13-15}
$n$ & $E^*$ & nodes & calls & $t_{E^*}$ & runtime & PDI & nodes & calls & $t_{E^*}$ & runtime & PDI & nodes & calls & runtime\\
 & & & & (s) & (s) & (s) & & & (s) & (s) & (s) & & & (s)\\
\midrule
16 & 24 & 61 & 1 & 0.03 & 0.03 & 0.03 & 61 & 1 & 0.03 & 0.03 & 0.03 & 61 & 1 & 0.07\\
17 & 32 & 319 & 13 & 0.03 & 0.03 & 0.03 & 319 & 13 & 0.03 & 0.04 & 0.03 & 319 & 13 & 0.15\\
18 & 25 & 1 & 1 & 0.00 & 0.00 & 0.00 & 1 & 1 & 0.00 & 0.00 & 0.00 & 1 & 1 & 0.05\\
19 & 29 & 411 & 73 & 0.03 & 0.05 & 0.03 & 411 & 73 & 0.03 & 0.06 & 0.04 & 411 & 73 & 1.04\\
20 & 26 & 390 & 200 & 0.07 & 0.09 & 0.06 & 328 & 130 & 0.03 & 0.09 & 0.05 & 328 & 130 & 2.25\\
21 & 26 & 439 & 186 & 0.07 & 0.08 & 0.05 & 381 & 129 & 0.03 & 0.09 & 0.05 & 381 & 129 & 2.32\\
22 & 39 & 1{,}968 & 915 & 0.14 & 0.32 & 0.10 & 1{,}777 & 730 & 0.03 & 0.41 & 0.11 & 1{,}777 & 730 & 13.71\\
23 & 47 & 14{,}604 & 7{,}519 & 0.03 & 2.36 & 0.78 & 14{,}604 & 7{,}519 & 0.03 & 3.98 & 1.30 & 14{,}604 & 7{,}519 & 148.21\\
24 & 36 & 12{,}782 & 7{,}203 & 2.35 & 2.34 & 0.87 & 4{,}457 & 2{,}466 & 0.03 & 1.51 & 0.56 & 4{,}388 & 2{,}425 & 55.82\\
25 & 36 & 12{,}788 & 8{,}253 & 2.69 & 2.69 & 0.95 & 5{,}692 & 2{,}953 & 0.03 & 1.80 & 0.53 & 5{,}522 & 2{,}853 & 70.20\\
26 & 45 & 6{,}704 & 3{,}202 & 0.03 & 1.11 & 0.22 & 6{,}810 & 3{,}255 & 0.03 & 2.07 & 0.39 & 6{,}631 & 3{,}166 & 82.89\\
\bottomrule
\end{tabular}
\end{adjustbox}
\caption{The same algorithm with SA at two smaller budgets per call (settings of
\cref{sec:setup}, seed~1; measured): nodes, sampler calls, time $t_{E^*}$ until the incumbent
first reaches $E^*$, runtime and PDI, with the $1{,}000$ shots $\times$ $1{,}000$ sweeps series
of \cref{tab:backends} for comparison. Every run proves the QOBLIB optimum $E^*$.}
\label{tab:sabudget}
\end{table}

The classical ingredients were measured in the same setting, the optimum supplied at the root
and no samples, so that only the classical part varies, by switching one of them off at a time
(\cref{tab:ablation}). Without propagation the tree is $1.9$--$2.6$ times larger, without the
reversal rule $1.9$--$2.0$ times larger, as $4{,}096$ class roots enter the queue instead of
$2{,}112$, without the pair term of the bound $23\%$ larger at $n=22$, $8\%$ at $n=26$ and
$0.3$--$1.3\%$ at $n=28$ and $30$, and with the lattice step $\Delta=4$ in place of $8$ it is
unchanged at $n=22$ and $26$ and $7\%$ larger at $n=30$ ($n=28$ has $\Delta=4$ in any case).

\begin{table}[tbp]
\centering\footnotesize
\begin{adjustbox}{max width=\textwidth}
\begin{tabular}{l rrrr}
\toprule
 & \multicolumn{4}{c}{nodes}\\
\cmidrule(lr){2-5}
 & $n=22$ & $n=26$ & $n=28$ & $n=30$\\
\midrule
full algorithm & 1{,}777 & 6{,}563 & 81{,}050 & 91{,}534\\
without propagation & 4{,}434 (+150\%) & 16{,}873 (+157\%) & 155{,}965 (+92\%) & 226{,}619 (+148\%)\\
without the pair term of the bound & 2{,}181 (+23\%) & 7{,}095 (+8.1\%) & 81{,}282 (+0.3\%) & 92{,}718 (+1.3\%)\\
lattice step $\Delta=4$ in place of $8$ & 1{,}777 (0\%) & 6{,}563 (0\%) & 81{,}050 (0\%) & 98{,}186 (+7.3\%)\\
without the reversal rule & 3{,}511 (+98\%) & 12{,}973 (+98\%) & 154{,}418 (+91\%) & 179{,}885 (+97\%)\\
\bottomrule
\end{tabular}
\end{adjustbox}
\caption{Node counts with one classical ingredient switched off, with the known optimum supplied
at the root and no samples ($\nu\equiv1$), so that only the classical part varies (seed~1;
measured; every run proves $E^*$). The first row is the full algorithm in this setting (the
``optimum at root'' series of \cref{tab:backends}). Without the pair term, the bound is the
per-lag bound $\LB_0$ of \cref{sec:branching} at every node, while the lag budgets of the
propagation rule keep the pair term. For $n=28$ the lattice step is $\Delta=4$ already, so that
row repeats the first.}
\label{tab:ablation}
\end{table}

Above the proven range, by contrast, the algorithm returns the incumbent with the certified bound
of \cref{thm:framework}(e) at its time limit, and the sampler matters most there, because the
incumbent is the reported answer and sets the gap. Memetic tabu search reaches best known values
far beyond $n=66$ \citep{boskovic2017labs}, as measured in that work, not here.

\subsection{Hardware resources and noise}
\label{sec:hardware}

The QAOA circuit of a node is specified by the monomials $w_SZ_S$ of $\Hnode$ and the angle
rescaling of \cref{sec:sampler}. On IBM Heron devices, fractional gates
execute $R_{ZZ}$ directly \citep{ibm2026fractional}; a noise-aware layout and routing is chosen by the
Qiskit preset transpiler at optimization level~3, dynamical decoupling (XY4) suppresses idle
errors, and measurement twirling symmetrizes the readout error. Pauli twirling of the two-qubit
gates is not supported with fractional gates \citep{ibm2026fractional}, and error mitigation by zero-noise extrapolation or
probabilistic error cancellation estimates expectation values rather than samples, so neither is
used. Quantinuum devices, in contrast, have all-to-all connectivity and a native arbitrary-angle
$R_{ZZ}$ gate \citep{moses2023racetrack,he2025iceberg}, so the four-spin terms need no routing. For error detection we use the
$[[k+2,k,2]]$ Iceberg code, which has been shown to improve QAOA samples on a trapped-ion device
\citep{he2025iceberg,jin2025iceberg}. The code has $k$ data qubits ($k=F$ rounded up to an even
number, an odd $F$ being padded with one idle logical qubit), a top qubit $t$ and a bottom qubit
$b$, with the stabilizers $X^{\otimes(k+2)}$ and $Z^{\otimes(k+2)}$ and the logical operators
$\bar X_i=X_tX_i$ and $\bar Z_i=Z_iZ_b$. A phase rotation $e^{-i\theta Z_S/2}$ of $\Hnode$ is
therefore the same physical rotation on $S$ when $|S|$ is even, and the rotation on $S\cup\{b\}$
when $|S|$ is odd, which is the case of the $1$- and $3$-body terms of a node Hamiltonian; the
mixer $e^{-i\beta X_i}$ is the physical rotation $R_{XX}(2\beta)$ on $t$ and $i$; and the logical
$|+\rangle^{\otimes k}$ is the $X$-basis GHZ state on all $k+2$ qubits. The master Hamiltonian
has only even terms, so its phase layer is the physical circuit of the unencoded one. In a
departure from \citep{he2025iceberg}, where syndrome measurements are interleaved with the gate
blocks, the syndrome is read once, at the end: the $X$ stabilizer through one ancilla and the $Z$
stabilizer as the parity of the final measurement of the $k+2$ qubits, so that errors that cancel
before the end pass undetected. Shots with a detected error are discarded. Two-qubit gate counts
are the numbers of native two-qubit gates after compilation to the device gate set; for the
encoded job they consist of the phase rotations, whose odd terms act on one more qubit than in
the unencoded circuit, the $F$ mixer gates $R_{XX}$ per layer, and the $k+1$ and $k+2$ CNOTs of
the encoding and of the syndrome readout.

\begin{figure}[tbp]
\centering
\includegraphics[width=\textwidth]{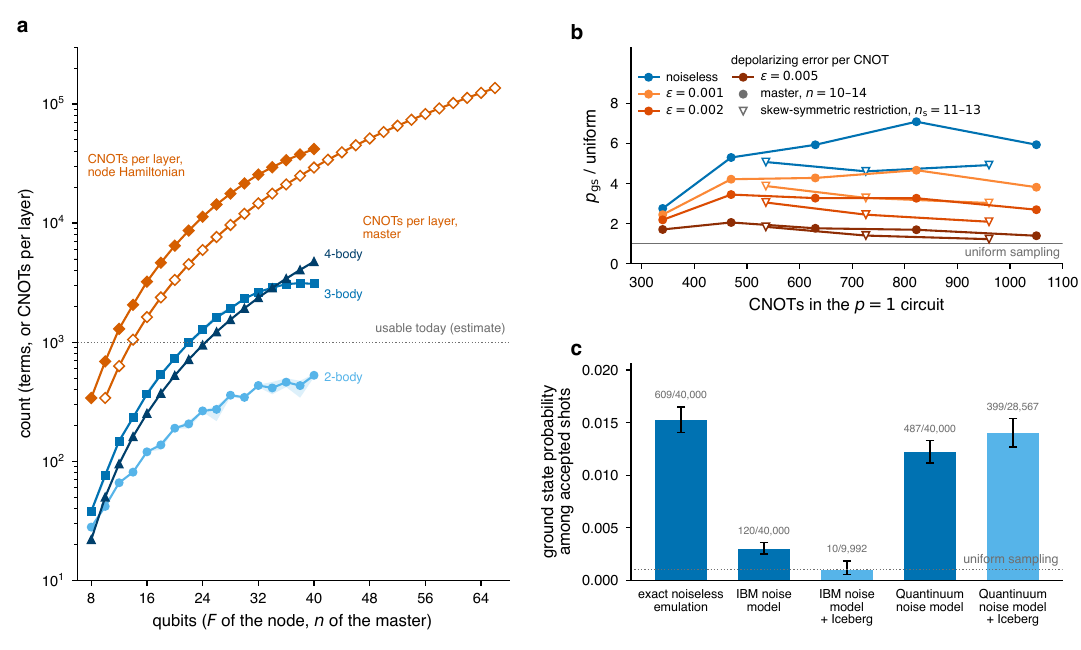}
\caption{Hardware resources and noise; every panel is an emulation or a model.
(a) Exact number of $2$-, $3$- and $4$-body monomials of
$\Hnode$ against the number of free spins $F$ (contiguous free block of an $n=50$ sequence,
random fixed boundary spins, mean and min--max band over $5$ draws; computed) and the CNOT count
of one QAOA phase layer in the ladder model $2\cdot\#ZZ+4\cdot\#ZZZ+6\cdot\#ZZZZ$ without routing
(model), for node Hamiltonians (filled) and the master on $n=10$--$66$ qubits (open). Dotted: the
budget of $10^3$ two-qubit gates, the circuit size that keeps a fidelity of $e^{-1.4}$ to
$e^{-3.2}$ at the two-qubit error rates of the H2-1 specification and of the Marrakesh
calibration snapshot (estimate). (b) Depth $p=1$ with a
depolarizing error $\epsilon$ on every CNOT (Qiskit Aer, $2{,}000$ trajectories, $1{,}000$ at
$\epsilon=0.005$; no other error): ground state probability relative to uniform
sampling against the CNOT count, master at $n=10$--$14$ (filled) and skew-symmetric restriction
at $n_{\mathrm s}=11$--$13$ (open). (c) Local emulation of one $p=1$ job on the class root of
$n=24$ with $F=10$ free spins whose ground state is the optimum, $40{,}000$ shots, seed~3, on the
emulators named (measured): ground state probability among accepted shots with Wilson $95\%$
intervals and the counts (ground states / accepted shots) above the bars, on the exact noiseless
emulation ($609/40{,}000$), the Qiskit noise model of the IBM Marrakesh device ($917$
two-qubit gates, estimated circuit fidelity $0.07$; $120/40{,}000$), the same with the Iceberg
code ($1{,}702$ gates, fidelity $0.004$, $25\%$ of the shots accepted; $10/9{,}992$), a
depolarizing model of a Quantinuum device ($279$ gates, fidelity $0.74$; $487/40{,}000$) and the
same with Iceberg ($450$ gates, fidelity $0.62$, $71\%$ accepted; $399/28{,}567$); light bars:
with Iceberg; dotted: uniform sampling ($0.0010$).}
\label{fig:hardware}
\end{figure}

\Cref{fig:hardware}(c) shows the emulated effect of noise at the size where present hardware is
usable: one $p=1$ job on the class root of $n=24$ with $F=10$ free spins whose ground state is
the optimum, $40{,}000$ shots with a single fixed seed (seed~3; no other seed was run), on five
emulators: the exact noiseless emulation, the Qiskit noise
model of the IBM Marrakesh device with its calibration data (calibration snapshot of 2025-02-26
\citep{qiskitibmruntime2026}; without fractional gates), and a depolarizing model of a Quantinuum
device, the latter two with and without Iceberg. The depolarizing model has $10^{-3}$ two-qubit,
$3\times10^{-5}$ single-qubit and $10^{-3}$ readout error rates, the H2-1 specification values
$1.36\times10^{-3}$, $3.4\times10^{-5}$ and $1.5\times10^{-3}$ \citep{quantinuum2025specs}
rounded down to one digit. The ground state
probability among the accepted shots is reported with its Wilson $95\%$ interval. The exact
noiseless emulation gives $0.0152$ ($0.0141$--$0.0165$; $609$ of $40{,}000$ shots), $15\times$
the uniform value $0.0010$. On heavy-hex connectivity \citep{ibm2026processors} the IBM model keeps $3\times$ uniform
without the code ($0.0030$, $0.0025$--$0.0036$; $917$ two-qubit gates) but falls to the uniform
level with it ($0.0010$, $0.0005$--$0.0018$; $1{,}702$ gates, $25\%$ of the shots accepted),
since Iceberg nearly doubles the two-qubit gates, so the code brings no gain in the IBM model. On
all-to-all connectivity, by contrast, the unencoded circuit keeps an estimated fidelity of $0.74$
(the product of the gate and readout fidelities of the model over the compiled circuit)
and gives $0.0122$ ($0.0112$--$0.0133$), while with Iceberg the probability among the accepted
shots is $0.0140$ ($0.0127$--$0.0154$) at $71\%$ of the shots accepted: post-selection removes
slightly more bad shots than good ones, a gain of $15\%$ (95\% interval $1$--$31\%$). A
depolarizing model cannot show what the code detects on a device (leakage, memory and correlated
errors); for that, the device results cited above are the evidence. By
\cref{thm:framework} none of this affects the validity of a proof.

\Cref{fig:hardware}(a) shows the exact term counts of $\Hnode$ as a function of the number $F$ of
free spins (for a contiguous free block of an $n=50$ sequence with random fixed boundary spins)
and the CNOT count
per QAOA layer. The number of $4$-body terms is the number of quadruples $a<b<c<d$ with
$b-a=d-c$, about $F^3/12$ (the figure $n^3/6$ often quoted counts $(k,i,j)$ triples, two of
which collapse into one monomial with coefficient~4), while the $3$-body terms scale like
$F^3/12$--$F^3/6$ depending on the fixed context and the $2$-body terms like $F^2/4$--$F^2/2$. A
QAOA layer with a CNOT ladder costs $6$, $4$, $2$ CNOTs per $ZZZZ$, $ZZZ$, $ZZ$ term (model, no
routing), which amounts to about $6{,}500$ CNOTs at $F=20$ and $21{,}000$--$23{,}000$ at $F=30$,
with depth $\sim F^2$ two-qubit layers on all-to-all connectivity; in the job of
\cref{fig:hardware}(c), routing and the CZ decomposition on the heavy-hex layout multiplied the
two-qubit gate count by $3.3$ ($917$ against $279$). With the two-qubit error rate
$1.36\times10^{-3}$ of the H2-1 specification \citep{quantinuum2025specs} and the median CZ error
$3.2\times10^{-3}$ of the Marrakesh calibration snapshot \citep{qiskitibmruntime2026}, a circuit
of $10^3$ two-qubit gates keeps a fidelity of $e^{-1.4}$ to $e^{-3.2}$ if the circuit fidelity is
estimated as the product of the two-qubit gate fidelities, so we take $10^3$ gates
as the upper limit of the usable budget (estimate).
This budget admits $p=1$ only at $F\le12$--$14$ for node Hamiltonians, $n\le18$--$20$ for the
master and $n_{\mathrm s}\le14$--$16$ for the skew-symmetric restriction, and at these sizes the
sampler gains little: in the trapped-ion runs of \citep{shaydulin2024qaoa} up to $n=18$, $p=1$
QAOA approaches random guessing at the largest sizes, and in a noise model with depolarizing
error $2\times10^{-3}$ per CNOT (Qiskit Aer, no single-qubit, readout or routing error), our
$p=1$ circuits keep a ground state probability of $2.1$--$3.5$ times the uniform value for the
master at $n=10$--$14$ and the skew-symmetric restriction at $n_{\mathrm s}=11$--$13$, against
$2.8$--$7.1$ times without noise (\cref{fig:hardware}(b)). So $p=1$ gains only a small factor
over uniform sampling at these sizes, even without noise, and a device run at these sizes would test
the execution of the circuits rather than the usefulness of the sampler. A proof, however, also
calls the sampler on many small Hamiltonians, because deep nodes have few free spins and fit the
budget. The master Hamiltonians of the instances proven here (up to $30$ qubits) and the class
roots at the largest $n$ ($F=16$ free spins at $n=30$) are beyond present hardware at useful
fidelity.

\section{Discussion and conclusion}
\label{sec:discussion}

\Cref{tab:components} lists each QCBB component and its LABS version. The node loop is that of
QCBB: lowest bound first, one sampler call per node, branching by conflict values from the node's
own samples, and both bounds updated and recorded at every node. The main changes are on the
classical side: the master Hamiltonian is the objective itself, so no penalty is needed; the
bound is an exact integer bound instead of an approximation-based one; propagation works on
dynamic lag budgets rather than on linear rows, and the energy lattice sharpens the pruning test;
and symmetry, which QCBB does not use, gives $4\times$ fewer leaves.

\begin{table}[t]
\centering\small
\begin{adjustbox}{max width=\textwidth}
\begin{tabular}{p{0.24\textwidth}p{0.32\textwidth}p{0.34\textwidth}}
\toprule
QCBB component & LABS version & Change and reason\\
\midrule
Master problem: BLP with $Ax=b$ & unconstrained quartic polynomial $\E(s)$ & linearization needs $\sim n^3/12$ binaries, $M\sim n^3/3$, GW slack $\gg E^*$\\
Master Hamiltonian: 2-local Ising, big $M$ & $H=\sum_k C_k^2$, 4-local, $n$ qubits, no $M$ & exact and penalty-free\\
Constraints for $V,\nu,P$ & lag budgets $C_k^2\le b_k$, $b_k=T-\LB_{-k}$ & implied by ``improve the incumbent''; feed eqs.~(5)--(8), propagation, infeasibility\\
Sec.~3.1 samples & fixed-angle QAOA on $\Hnode$ at every node, $1{,}000$ shots & no variational loop; exact energies, $V$, $\nu$\\
Eqs.~(5)--(9) conflict values & eq.~(7) by analogy; two outermost free spins as candidates & equal to eq.~(9) on nodes with a fixed prefix and suffix (\cref{prop:frontier})\\
Eqs.~(10)--(16) subproblem & substitution into the polynomial, constant tracked & polynomial instead of linear rows\\
Sec.~3.2.4 propagation & budget propagation (\cref{thm:prop}) & forcing of spins\\
Sec.~3.3 node selection & lowest bound first; ties: deeper, then older & unchanged\\
Eqs.~(17), (19) bound & ranges $R_k$ plus pair congruence (\cref{thm:perlag,thm:pair}); integer; monotone & GW bound holds only in expectation and far too loose\\
Sec.~3.5 pruning & $\LB>\UB-\Delta$ & lattice step (\cref{lem:lattice})\\
Sec.~3.5.1, App.~C & empty budget set, propagation contradiction & same role, no $M$\\
Sec.~3.6 stopping, gap & lowest open bound $>T$; certified $\LB$ on the lattice (\cref{thm:framework}) & bounds recorded against time\\
symmetry (not in QCBB) & group of order 8, $s_1=s_2=+1$; reversal rule on class roots & $4\times$ fewer leaves; master branches into class roots\\
\bottomrule
\end{tabular}
\end{adjustbox}
\caption{QCBB components and their LABS versions.}
\label{tab:components}
\end{table}

On the quantum side, in the proven regime the samples shorten a proof only by supplying the
optimal $\UB$ early; once the incumbent is optimal, their branching choice enlarges the tree
against the sample-free rule $\nu\equiv1$, by $7$--$15\%$ at $n=24$--$26$ and by $48$--$76\%$ at
$n=27$, $28$ and $30$ for uniform samples with the optimum supplied at the root, and with the optimum at the
root QAOA at $p=12$ gives the same tree as uniform samples at $n=24$ and nearly the same at $n=27$
($12{,}433$ against $12{,}424$ nodes; \cref{prop:samples},
\cref{sec:samples}). This is a property of exact branch and bound with
sample-guided branching rather than of LABS: \citep{hadfield2026iqo} reaches the same structure for
a tree search over QAOA-guided reductions on MaxCut, where quantum output ranks branches and
supplies candidates while pruning is justified classically, and it reports that a classical
greedy rule often matches or improves on the tested QAOA ordering. The sampler
matters most beyond the proven range, where its incumbent is the reported answer and sets the
gap.

The angle rescaling and the skew-symmetric restriction raise the ground state probability per shot in
our measurements (\cref{sec:sampler}, \cref{fig:qaoa}). Inside the full algorithm, QAOA at $p=12$
reaches $E^*$ with the first call at $n\le26$, whereas QAOA at $p=1$ misses $E^*$ more often and
needs up to $2.9\times$ more nodes. The number of calls and shots is the primary cost measure;
the runtime and the PDI, in turn, follow the cost of a call in the simulation, which for QAOA
grows like $2^F$ with the size of the node Hamiltonian ($5.0$\,s on the master at $n=26$) and is
$1.0$--$2.9$\,ms on average over the nodes of a proof. That cost is a property of how the
circuit is executed, not of the sampler, and this is what the PDI is for. Any sampler enters the
same two metrics, with its execution time counted as sampler time. Two quantum optimization
approaches, or a quantum and a classical one, can thus be compared by a proof or a certified gap
and by a convergence measure on one scale, rather than by the time to reach an optimum that is
already known. The optima proved here were known from \citep{packebusch2016labs};
what is new is their proof by a quantum optimization algorithm, with a certificate and with a
measured convergence.

The approach has the following limitations. The proven range, $n\le30$, is far below the $n=66$
of the classical branch and bound of \citep{packebusch2016labs} and the $n=41$ of Gurobi
\citep{koch2026qoblib}, because one sampler call per node costs milliseconds, against
microseconds per node in the classical branch and bound, and because the node count grows by
about $1.9$ per unit of $n$ in our fit over $n=20$--$30$, whose $95\%$ interval $1.59$--$2.37$
does not separate it from plain enumeration (\cref{sec:experiments}). The cost of our QAOA calls grows like
$2^F$ (\cref{sec:setup}), and a proof at $n=28$--$30$ makes about $10^5$ sampler calls. The
master Hamiltonians of the proven instances and the class roots
at the largest $n$ are beyond present hardware at useful fidelity (\cref{sec:hardware}). The
angle scale $\lambda$ was chosen on class roots whose ground state is the optimum, that is, with
knowledge of $E^*$, and then frozen, and at $p=12$ it sits beside the peak of the measured
geometric mean (\cref{sec:sampler}); the validity of a proof does not depend on it, while the
time to $E^*$ and, through the branching choice, the node count and the PDI may. The
bound, the propagation rule and the symmetry rule are specific to LABS, so what transfers to
another problem is \cref{thm:framework} and the pattern of the lag budgets, node-specific
constraints implied by ``improve the incumbent''. The angle rescaling is a conjecture.

The following conjecture is supported by measurements only and is not used on the guaranteed
path. The growth of the node count is not stated as a conjecture, because the fit over
$n=20$--$30$ does not determine it (\cref{sec:experiments}).

\begin{conjecture}[angle scale]
For node Hamiltonians whose ground state is the global optimum, the fixed angles transfer with
$\gamma_l=\lambda\,\bar\gamma_l/n$ and $\lambda\approx0.56$--$0.63$ for $p\ge2$ (the grid values
$0.5625$ and $0.625$ of \cref{sec:sampler}; $\lambda\approx0.875$ at
$p=1$), and with $\lambda\approx0.5$ for the skew-symmetric restriction (noiseless, $F\le28$
free spins, $n_{\mathrm s}\le26$).
\end{conjecture}

Open questions on the heuristic side are whether any sample-driven branching rule beats the
classical order from both ends inward on LABS, and whether a quantum sampler can return the incumbent at sizes where the master
and the class root Hamiltonians have $F\ge30$.

In conclusion, a quantum optimization algorithm, fixed-angle QAOA inside an exact
quantum-classical hybrid branch and bound, proved the optima of all lengths $n=2$--$30$ at $p=12$
and at $p=1$ (\cref{thm:framework,cor:labs}), and the bounds recorded against time and the
primal-dual integral measure the convergence of each run. The optima were known; what is new is
their
proof by a quantum optimization algorithm, with a certificate and a convergence measure that the
existing quantum approaches to LABS do not give. We claim no speedup.

\section*{Code and data availability}

The code of the solver (Python 3.13.7 with NumPy 2.5.4, compiled with Numba 0.68.1), with the
QAOA sampler, the SA sampler and the noise emulations with Qiskit Aer, is available at
\url{https://github.com/bentaxplus/qcbb-labs} under the MIT license. The run logs behind every
number in this paper and the scripts that produce every figure and table from them are available
from the author on request, and to the editors and referees of this paper.
The instances are labs002--labs100 of QOBLIB \citep{koch2026qoblib}. Our QOBLIB submission contains, for
labs002--labs030, the sequences, run summaries and incumbent time series of the five runs per
instance of QAOA at $p=12$ (\cref{app:seeds}); the optima themselves were already listed there.

\section*{Author contributions}

The author designed the algorithm, wrote the proofs and the code, ran the measurements and wrote
the paper. AI assisted in coding and in formulating the manuscript: Claude Code (Anthropic) with
the model Claude Opus 5.5 assisted in writing the Python code, the experiment scripts and the
text, and Claude Fable 5.1 assisted in challenging ideas and in verifying them. The author
directed the work, checked the code, the proofs and the results, and is responsible for the
content.

\bibliographystyle{quantum}
\bibliography{refs}

\appendix

\section{Correctness of the framework}
\label{app:framework}

This appendix states and proves the correctness theorem of \cref{sec:correctness}, and records
its two instances: QCBB's original binary linear program and LABS. The theorem refers to a lower
bound $\LB(N)$ with $\LB(N)\le f(x)$ for every $x\in\comp(N)$; to a propagation rule that maps a
node with threshold $T$ either to ``infeasible'' or to a node with more fixed spins; and to a
symmetry rule for a group under which $f$ is invariant.

\begin{theorem}[correctness of the framework]\label{thm:framework}
Assume
\begin{enumerate}
\item[(i)] the bound is valid: $\LB(N)\le f(x)$ for every node $N$ and every $x\in\comp(N)$;
\item[(ii)] propagation removes only completions with $f>T$, and returns ``infeasible'' only when
no completion has $f\le T$;
\item[(iii)] the symmetry rule discards a node only if no completion lies in $K$, where $K$ meets
every orbit of a group under which $f$ is invariant;
\item[(iv)] $f(X)\subseteq f_0+\Delta\mathbb{Z}$ and all thresholds are taken on this lattice;
\item[(v)] the children of a node partition its completions.
\end{enumerate}
Then, for every sampler:
\begin{enumerate}
\item[(a)] the search tree is finite and each node is evaluated once;
\item[(b)] if some $x$ has $f(x)\le T$, the search at threshold $T$ reaches a leaf with $f\le T$
(a search that does not stop at its first leaf reaches an element $x'\in K$ of the orbit of $x$,
$f(x')=f(x)$);
\item[(c)] if the search at $T=\UB-\Delta$ ends with no leaf of value $\le T$, $\UB=\min f$;
\item[(d)] if a search at a lattice threshold $T'$ ends with no leaf of value $\le T'$,
$\min f\ge T'+\Delta$;
\item[(e)] during a search at $T$, let $O$ be the set of open nodes: nodes created (all children
of a branched node count as created) and not yet closed, where a node is closed once it is
pruned, found infeasible, discarded by the symmetry rule, a leaf, or has all children closed; a
node not yet evaluated carries its parent's bound or any valid bound not below it (in
\cref{alg:main}, the key $\max(\LB(N),\LB(C))$). Then
$\min f\ge\min(T+\Delta,\min_{N\in O}\LB(N))$;
\item[(f)] $\LB_{\mathrm{cert}}=\max(\text{highest completed }T'+\Delta,\ \text{bound (e)})$
satisfies $\LB_{\mathrm{cert}}\le\min f\le\UB$, so $(\UB-\LB_{\mathrm{cert}})/\UB$ is a valid gap.
\end{enumerate}
Sample data enter only as incumbent candidates (re-evaluated exactly) and as the branching
variable; none of (a)--(f) depends on them.
\end{theorem}

\begin{proof}[Proof of \cref{thm:framework}]
(a) Every branching fixes at least one variable and propagation only fixes variables, so the
depth is at most $n$; each node has finitely many children. (b) Induction on depth. Let $x$ have
$f(x)\le T$ and let $x'\in K$ lie in the orbit of $x$; $f(x')=f(x)\le T$ by invariance. The root
contains $x'$. If a node $N$ contains $x'$, it is not pruned by the bound (by (i),
$\LB(N)\le f(x')\le T$), not removed by propagation ((ii): $x'$ has $f\le T$, so $x'\in\comp(N')$
and $N'$ is not infeasible), not discarded by the symmetry rule ((iii): $x'\in\comp(N)\cap K$),
and by (v) exactly one child contains $x'$. So $x'$ is a completion of a surviving node at every
depth, hence of a leaf. (c) $\UB$ is attained. If some $x$ had $f(x)<\UB$ then $f(x)\le\UB-\Delta=T$
by (iv), and (b) would give a leaf of value $\le T$. (d) By (b) no $x$ has $f(x)\le T'$, and by
(iv) the next lattice value is $T'+\Delta$. (e) Let $x$ be arbitrary. If $f(x)>T$, then
$f(x)\ge T+\Delta$ by (iv) and the claim holds for $x$. So let $f(x)\le T$ and let $x'$ be as in
(b); we show that $x'$ is a completion of an open node $N$, which gives
$f(x)=f(x')\ge\LB(N)\ge\min_{N\in O}\LB(N)$ by (i) (for an unevaluated node the parent's bound,
or any valid bound not below it, is valid because its completions are a subset of the parent's).
Suppose, for a contradiction, that every node containing $x'$ is closed. Follow $x'$ from the
root: by (iii) and (v) it passes from a closed node to a closed child until it reaches a node
that is pruned, found infeasible or a leaf (it is not discarded, since $x'\in K$, or $x'\in K'$
when the reversal rule is used, \cref{cor:labs}). A node pruned or found infeasible at an
earlier threshold $T_0\ge T$ (thresholds only decrease) has no completion with $f\le T_0$ by (i)
and (ii), which contradicts $f(x')\le T\le T_0$, and a leaf already seen has value $\ge\UB>T$,
which contradicts $f(x')\le T$. Hence $x'$ lies in an open node.
(f) follows from (d), (e) and $\UB=f(x_{\UB})$. The last sentence holds because (i)--(v) do not
refer to samples, (b) holds for every choice of branching variable and every node order, and (c)--(f)
hold for every attained $\UB$.
\end{proof}

For QCBB's original binary linear program with integer objective, the instance of
\cref{thm:framework} is: the MaxCut bound of QCBB eq.~(17) as (i), in its certified form with the
semidefinite value in place of the Goemans--Williamson value, which holds in expectation (QCBB
App.~B, \citealp{goemans1995maxcut}); variable bound propagation (Sec.~3.2.4) and the
infeasibility test (App.~C) as (ii); the trivial group as (iii); $\Delta=1$ as (iv); binary
branching as (v). We make no new claim about that instance.

\begin{corollary}[LABS satisfies \cref{thm:framework}]\label{cor:labs}
With the bound of \cref{thm:pair} (i), the propagation of \cref{thm:prop} (ii), the restriction
$s_1=s_2=+1$ of \cref{thm:sym}(a) and the reversal rule of \cref{thm:sym}(b) (iii), the lattice of
\cref{lem:lattice} (iv), and as (v) binary branching on any free spin below the class roots and,
at the master, the $2^{2m_0}$ assignments of the boundary spins, of which those with $s_1=s_2=+1$
are the class roots and the others are discarded by \cref{thm:sym}(a), the algorithm of
\cref{alg:main} returns a proven optimum or a valid gap for every sampler.
\end{corollary}
\begin{proof}
Conditions (i)--(v) are proved in the results named, and the conclusion is
\cref{thm:framework}(c)--(f).
\end{proof}

\section{Energy lattice}
\label{app:lattice}

This appendix proves that the LABS energies lie on a lattice of step $\Delta$, with $\Delta=8$
for $n\equiv 2 \pmod 4$ and $\Delta=4$ otherwise: condition (iv) of \cref{thm:framework}. The
argument uses that an odd square is $1$ modulo~$8$ and an even square is $0$ modulo~$4$.

\begin{lemma}[parity]\label{lem:parity}
$C_k\equiv n-k \pmod 2$.
\end{lemma}
\begin{proof}
$C_k$ is a sum of $n-k$ terms in $\{+1,-1\}$. A sum of $m$ such terms with $t$ terms equal to $-1$
equals $m-2t\equiv m \pmod 2$.
\end{proof}

\begin{lemma}[pair congruence; {\citealp{wiggenbrock2010labs}}, {\citealp[Sec.~3.1.3, eq.~(15)]{packebusch2016labs}}]\label{lem:pair}
For $1\le k\le n-1$, $C_k+C_{n-k}\equiv n \pmod 4$.
\end{lemma}
The congruence is attributed to \citep{wiggenbrock2010labs} and proved in
\citep{packebusch2016labs} by a spin flip argument. For $k=n/2$ the statement
is \cref{lem:parity} for that lag.

\begin{lemma}[energy residue]\label{lem:residue}
$\E\equiv\lfloor n/2\rfloor \pmod 4$.
\end{lemma}
\begin{proof}
An odd square is $1 \pmod 8$, an even square is $0 \pmod 4$. By \cref{lem:parity}, $C_k$ is odd iff
$n-k$ is odd. The number of $k\in\{1,\dots,n-1\}$ with $n-k$ odd is the number of odd integers in
$\{1,\dots,n-1\}$, which is $\lfloor n/2\rfloor$. Hence $\E=\sum_k C_k^2\equiv\lfloor n/2\rfloor
\pmod 4$.
\end{proof}

\begin{lemma}[lattice step]\label{lem:lattice}
\begin{enumerate}
\item[(a)] For every $n$, all attainable energies lie in one residue class modulo $4$.
\item[(b)] For $n=4m+2$, all attainable energies lie in one residue class modulo $8$, namely
$\E\equiv 6m+1 \pmod 8$.
\item[(c)] Hence $\E(\{-1,+1\}^n)\subseteq f_0+\Delta\mathbb{Z}$ for some $f_0$: condition (iv) of
\cref{thm:framework}. In particular, if $\UB=\E(s_{\UB})$ is attained and $\E(s)<\UB$, then
$\E(s)\le\UB-\Delta$.
\end{enumerate}
\end{lemma}
\begin{proof}
(a) is \cref{lem:residue}. (b) Pair the lags $(k,n-k)$ for $k=1,\dots,2m$ and keep the middle lag
$n/2=2m+1$. For odd $k$ ($m$ pairs), $n-k$ is odd, so $C_k$ and $C_{n-k}$ are odd
(\cref{lem:parity}), their squares are $1 \pmod 8$, and the pair contributes $2 \pmod 8$. For even
$k$ ($m$ pairs) both are even; write $a=2a'$, $b=2b'$. By \cref{lem:pair}, $a+b\equiv 2 \pmod 4$,
so $a'+b'$ is odd, exactly one of $a',b'$ is odd, $a'^2+b'^2$ is odd, and
$a^2+b^2=4(a'^2+b'^2)\equiv 4 \pmod 8$. For the middle lag, $n-(2m+1)=2m+1$ is odd, so $C_{n/2}$ is
odd and its square is $1 \pmod 8$. Altogether $\E\equiv 2m+4m+1=6m+1 \pmod 8$. (c) follows.
\end{proof}

\section{Bound}
\label{app:bounds}

This appendix defines the node Hamiltonian, the lag chains and free chains, and proves the exact
ranges $R_k$ and the pair congruence bound of \cref{sec:branching,sec:bounds}: condition (i) of
\cref{thm:framework}.

For a node $(\Phi,U)$ and lag $k$, substituting the fixed spins (QCBB, eqs.~(10)--(16)) gives
\begin{equation}
\begin{gathered}
  C_k(s)=c_k+\sum_{i\in U} h_{ki}\, \hat s_i+\sum_{i,\,i+k\in U} \hat s_i \hat s_{i+k},\\
  c_k=\sum_{i,\,i+k\in\Phi} s_i s_{i+k},\qquad
  h_{ki}=[i-k\in\Phi]\,s_{i-k}+[i+k\in\Phi]\,s_{i+k},
\end{gathered}
  \label{eq:sub}
\end{equation}
where the free spins $\hat s_i$ are indexed by their positions $i\in U$. The node Hamiltonian is
\begin{equation*}
  \Hnode(\hat s)=\sum_k C_k(\hat s)^2=\E(\text{completion}),
\end{equation*}
and collecting monomials ($\hat s_i^2=1$) yields the reduced Hamiltonian and its constant, the
analog of QCBB eq.~(16).

\begin{figure}[tbp]
\centering
\begin{adjustbox}{max width=\textwidth}
\definecolor{qblue}{HTML}{0072B2}
\definecolor{qverm}{HTML}{D55E00}
\definecolor{qgreen}{HTML}{009E73}
\definecolor{qpurple}{HTML}{CC79A7}
\begin{tikzpicture}[x=1mm, y=1mm, font=\scriptsize,
  spin/.style={circle, draw, line width=0.5pt, inner sep=0pt, minimum size=3.6mm},
  fixed/.style={spin, fill=black!80, text=white, font=\scriptsize\bfseries},
  free/.style={spin, fill=white},
  edge/.style={line width=0.8pt},
  brace/.style={decorate, decoration={brace, amplitude=1.2mm}, line width=0.5pt},
]
  \def\dx{10}
  \foreach \i in {0,...,12} {
    \pgfmathsetmacro{\x}{\i*\dx}
    \ifnum\i=0 \node[fixed] (s\i) at (\x,0) {$+$}; \else
    \ifnum\i=6 \node[fixed] (s\i) at (\x,0) {$+$}; \else
    \ifnum\i=8 \node[fixed] (s\i) at (\x,0) {$+$}; \else
    \node[free] (s\i) at (\x,0) {}; \fi\fi\fi
    \node[below=0.3mm of s\i, font=\tiny] {$s_{\i}$};
  }
  \draw[edge, qblue]   (s0) to[bend left=60] (s2) (s2) to[bend left=60] (s4) (s4) to[bend left=60] (s6);
  \draw[edge, black!50] (s6) to[bend left=60] (s8);
  \draw[edge, qgreen]  (s8) to[bend left=60] (s10) (s10) to[bend left=60] (s12);
  \draw[edge, qpurple] (s1) to[bend right=60] (s3) (s3) to[bend right=60] (s5) (s5) to[bend right=60] (s7)
                       (s7) to[bend right=60] (s9) (s9) to[bend right=60] (s11);
  \draw[brace, qblue] (-2,8.5) -- (6*\dx+2,8.5)
    node[midway, above=1.4mm, align=center, text=qblue] {type (i): two fixed ends, $e=3$, $s_0s_6=+1$\\ $S_1=\{-1,\,3\}$ (step 4)};
  \node[align=center, text=black!50] at (7*\dx, 14.5) {fixed product\\ $c_2=s_6s_8=+1$};
  \draw[brace, qgreen] (8*\dx-2,8.5) -- (12*\dx+2,8.5)
    node[midway, above=1.4mm, align=center, text=qgreen] {type (ii): one fixed end, $e=2$\\ $S_2=\{-2,0,2\}$ (step 2)};
  \draw[brace, qpurple] (11*\dx+2,-10) -- (\dx-2,-10)
    node[midway, below=1.4mm, align=center, text=qpurple] {chain $r=1$, type (iii): no fixed end, $e=5$ products, $6$ free spins, $S_3=\{-5,-3,\dots,5\}$ (step 2)};
  \node[anchor=east, text=black!60] at (-5,4.5) {chain $r=0$};
  \node[anchor=east, text=black!60] at (-5,-4.5) {chain $r=1$};
  \node[anchor=west, align=left] at (0,-22)
    {$R_2 \;=\; c_2+S_1+S_2+S_3 \;=\; 1+\{-1,3\}+\{-2,0,2\}+\{-5,\dots,5\} \;=\; \{-7,-5,\dots,11\}$\\[1pt]
     $\mu_2=\min_{v\in R_2}|v|=1$;\quad enumeration of all $2^{10}$ completions gives the same set};
  \begin{scope}[shift={(0,-36)}]
    \draw[-{Stealth[length=1.2mm]}, line width=0.5pt] (-2,0) -- (12*\dx+6,0) node[right, font=\tiny] {$C_2$};
    \foreach \v in {-8,-6,...,12} {
      \pgfmathsetmacro{\x}{(\v+8)*\dx/2}
      \draw[line width=0.4pt] (\x,-0.8) -- (\x,0.8);
      \node[below=0.6mm, font=\tiny] at (\x,0) {$\v$};
    }
    \foreach \v in {-7,-5,...,11} {
      \pgfmathsetmacro{\x}{(\v+8)*\dx/2}
      \fill[qverm] (\x,0) circle (0.9mm);
    }
    \draw[qverm, line width=0.5pt] (7*\dx/2, 3.5) -- (7*\dx/2, 1.3);
    \node[above, text=qverm, font=\tiny] at (7*\dx/2, 3.5) {$\mu_2=1$};
  \end{scope}
\end{tikzpicture}
\end{adjustbox}
\caption{Lag chains, free chain types and the range $R_k$ of \cref{thm:perlag} on a concrete
node: $n=13$, lag $k=2$, $s_0=s_6=s_8=+1$ fixed (filled) and the other ten spins free (open;
0-based positions). The lag-2 products form two chains (arcs above: even positions; below: odd
positions), and the fixed spins cut them into free chains: $s_0$--$s_6$ has two fixed ends (type (i);
$s_0s_6=+1$ forces an even number of $-1$ products, so $S_1=\{-1,3\}$, step 4), $s_8$--$s_{12}$ has
one fixed end (type (ii), $S_2=\{-2,0,2\}$), the odd chain has none (type (iii),
$S_3=\{-5,-3,\dots,5\}$), and the fixed product $s_6s_8$ contributes $c_2=+1$. Free chains share no
free spin, so $R_2=c_2+S_1+S_2+S_3=\{-7,-5,\dots,11\}$ (number line) and
$\mu_2=\min_{v\in R_2}|v|=1$. Checked against enumeration of all $2^{10}$ completions
(computed).}
\label{fig:chains}
\end{figure}
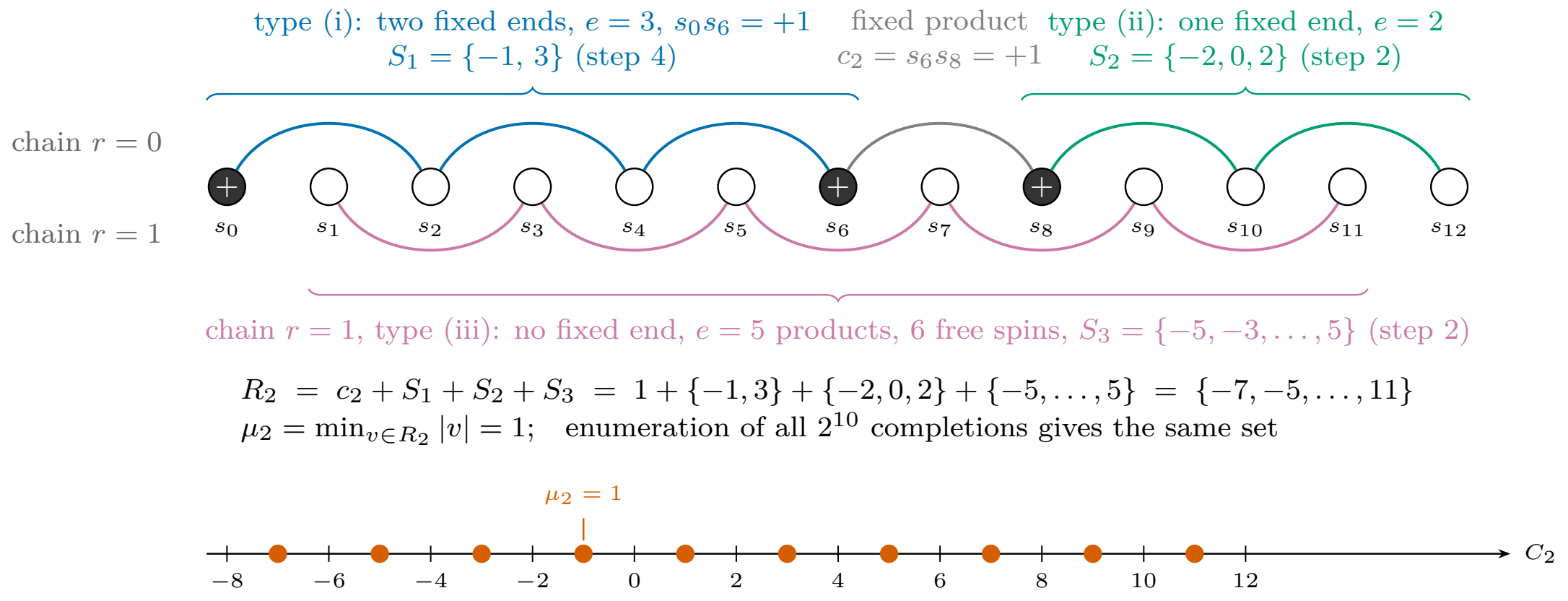

The range $R_k$ of $C_k$ is the set of values that $C_k$ takes over all completions of the node.
Its structure follows from a picture of the lag-$k$ products (\cref{fig:chains}). The positions
split into $k$ chains $r,r+k,r+2k,\dots$, and every lag-$k$ product joins two consecutive
positions of one chain (a chain of a single position has no product and plays no role). A product
is \emph{free} if at least one of its positions is free. Cutting every chain at its fixed
positions leaves \emph{free chains}, in the words of \citep{packebusch2016labs}: maximal runs of
consecutive free products with no fixed position in the interior. A free chain with $e$ products
is of type (i) if both its end positions are fixed, of type (ii) if exactly one end is fixed (the
run reaches the chain end), and of type (iii) if the whole chain is free. Free chains share no
free spin, so their sums vary independently. The step of $R_k$ (the granularity of
\citep{packebusch2016labs}) is $4$ if every free chain is of type (i), because the two fixed ends
fix the parity of the number of $-1$ products, and $2$ otherwise. This is the intuition behind
\cref{thm:perlag}: the range is a Minkowski sum of the ranges of the independent free chains, each
of which is an arithmetic progression.

\Cref{thm:perlag} is the ``tight bound'' of \citep[Sec.~3.1.5]{packebusch2016labs} for the
contiguous free block left by its shells. Their induction for the range of a chain with both ends
fixed (their eq.~(25)) is written out for odd run lengths; the even case is stated to be similar
and omitted. The plain range $c_k\pm f_k$ of step $2$, with $f_k$ the number of free lag-$k$
products, underlies their eq.~(12), stated there in their notation as $\max(b_k,|c_k|-f_k)$, and
\citep{prestwich2013labs}. Our proof covers both parities and arbitrary fixed sets, which is
needed because propagation fixes interior spins. The computed sets were checked against
enumeration at random nodes with arbitrary fixed sets (\cref{sec:setup}).

\begin{theorem}[per-lag ranges]\label{thm:perlag}
Fix a node $(\Phi,U)$ and a lag $k$. The set $R_k$ of values that $C_k$ takes over all completions
is an arithmetic progression
\[
  R_k=\{\lo_k,\ \lo_k+g_k,\ \dots,\ \hi_k\},\qquad g_k\in\{2,4\},
\]
where $\lo_k-c_k$ and $\hi_k-c_k$ are the sums of the minima and of the maxima of the free chain
sums, and $g_k=4$ if every free chain is of type (i) and $g_k=2$ otherwise. $R_k$ is computable in $O(n-k)$
integer operations. Consequently, with $\mu_k=\min\{|v|:v\in R_k\}$, every completion satisfies
$\E(s)\ge\LB_0:=\sum_k\mu_k^2$; $\LB_0$ does not decrease from a node to its children; and
$\LB_0\equiv\lfloor n/2\rfloor \pmod 4$.
\end{theorem}

\begin{proof}[Proof of \cref{thm:perlag}]
Write the positions as $k$ chains $r,r+k,r+2k,\dots$ ($r=0,\dots,k-1$, 0-based); every lag-$k$
product $s_a s_{a+k}$ is a product of two consecutive positions of one chain, and a chain of a
single position (possible for $k>(n-1)/2$) carries no product and is ignored. Call a product free
if at least one of its positions is in $U$, fixed otherwise, and let $c_k$ be the sum of the fixed
products. Cut every chain at each of its fixed positions. A \emph{free chain} is a piece between two
consecutive fixed positions of a chain, or between a chain end and the nearest fixed position, or
a whole chain without fixed positions, that contains at least one free position, together with
the products along it; equivalently, a maximal run of consecutive free products with no fixed
position in its interior. Products between two adjacent fixed positions are fixed. Every interior
position of a free chain is free; only its end positions can be fixed. Two free chains share no
free position: a free position $a$ lies in the products $(a-k,a)$ and $(a,a+k)$, which are
consecutive with no fixed position between them, hence in the same free chain. Therefore, as the
free spins range over $\{-1,1\}^U$, the free chain sums vary independently, and
$R_k=c_k+(S_1+\dots+S_J)$ is a Minkowski sum of the sets $S_j$ of sums attainable by the $J$ free
chains alone. For one free chain with $e\ge1$ products (\cref{fig:chains}), the three cases below
are exhaustive because its interior positions are free (a type-(i) free chain has $e\ge2$, since
it contains a free position):
\begin{enumerate}
\item[(i)] Both end positions fixed (at $u$ and $w$), $e-1$ free spins between them. The map from
the free spins to the product vector $p\in\{-1,1\}^e$ is injective with image
$\{p:\prod_i p_i=s_u s_w\}$: given $p$, the spins are $s_{u+ik}=s_u p_1\cdots p_i$, and the last
product equals $s_u s_w$ iff $\prod p=s_u s_w$. The sum is $e-2t$ with $t$ the number of $-1$
entries, $t\equiv[s_us_w=-1] \pmod 2$, $0\le t\le e$: an arithmetic progression of step $4$ from
$-e$ or $-(e-2)$ to $e$ or $e-2$ as the parity allows.
\item[(ii)] Exactly one end fixed (the run reaches the chain end), $e$ free spins: spins and
product vectors are in bijection; $S_j=\{-e,-e+2,\dots,e\}$, step $2$.
\item[(iii)] No end fixed (the whole chain is free), $e$ products and $e+1$ free spins: every
product vector is attained (twice); $S_j=\{-e,\dots,e\}$, step $2$.
\end{enumerate}
A Minkowski sum of step-$4$ progressions is a step-$4$ progression from the sum of the minima to
the sum of the maxima. If at least one summand is a step-$2$ progression $\{a,a+2,\dots\}$ with at
least two elements (every type-(ii) or (iii) free chain, as $e\ge1$), the sum is the full step-$2$
progression between the sum of the minima and the
sum of the maxima: for a target $x$ of the right parity, let $z$ be the largest element of the
remaining sum with $z\le x-a$ (or its top if $x-a$ exceeds it); then $x-z\in[a,a+4)$ with the
right parity, i.e.\ $x-z\in\{a,a+2\}$, which lies in the step-$2$ summand (in the top case
$x-z$ is at most the summand's maximum and at least $a$). Hence $g_k=4$ if every free chain is of
type (i) and $g_k=2$ otherwise, and $\lo_k$, $\hi_k$ are $c_k$ plus the sums of the free chain
minima and maxima. Every element of $R_k$ has the parity of $n-k$ (\cref{lem:parity}). The chain scan
costs $O(n-k)$.

For the bound, $C_k(s)\in R_k$ for every completion, so $C_k(s)^2\ge\mu_k^2$ and $\E(s)\ge\LB_0$.
$\mu_k$ is read off in $O(1)$: it is $\lo_k$ if $\lo_k\ge0$, $-\hi_k$ if $\hi_k\le0$, and otherwise
$\min(x,g_k-x)$ with $x$ the smallest non-negative element of the progression. A child's
completions are a subset of the parent's, so $R_k$ shrinks and $\mu_k$ does not decrease. Finally
$\mu_k\equiv n-k \pmod 2$, so $\sum_k\mu_k^2\equiv\lfloor n/2\rfloor \pmod 4$ by the argument of
\cref{lem:residue}.
\end{proof}

The pair congruence bound of \cref{sec:bounds} is
\begin{equation}
  \begin{gathered}
  \LB=\sum_{k<n/2} m_k+[n\text{ even}]\,\mu_{n/2}^2,\\
  m_k=\min\{a^2+b^2:\ a\in R_k,\ b\in R_{n-k},\ a+b\equiv n \pmod 4\}.
  \end{gathered}
  \label{eq:lb}
\end{equation}

\begin{theorem}[bound with the pair congruence]\label{thm:pair}
For $k<n/2$ let
\[
  m_k=\min\{a^2+b^2:\ a\in R_k,\ b\in R_{n-k},\ a+b\equiv n \pmod 4\}.
\]
Then every completion satisfies $\E(s)\ge\LB:=\sum_{k<n/2} m_k+[n\text{ even}]\,\mu_{n/2}^2\ \ge\ \LB_0$,
and each $m_k$ is computed in $O(1)$ from the two progressions. At the root, for $n\equiv2\pmod 4$,
$\LB=3n/2-2$. In particular $\LB$ satisfies condition (i).
\end{theorem}

\begin{proof}[Proof of \cref{thm:pair}]
For any completion, $(C_k,C_{n-k})\in R_k\times R_{n-k}$ by \cref{thm:perlag} and
$C_k+C_{n-k}\equiv n \pmod 4$ by \cref{lem:pair}, so the pair lies in the feasible set of the
minimum, which is therefore non-empty, and $C_k^2+C_{n-k}^2\ge m_k\ge\mu_k^2+\mu_{n-k}^2$. Summing
over pairs and adding the middle lag gives $\LB\ge\LB_0$. The objective is separable: for each
residue $v\in\{0,1,2,3\}$ the best $a$ is the element of $R_k$ with $a\equiv v \pmod 4$ and the
smallest $|a|$ (if any) and the best $b$ the element of $R_{n-k}$ with $b\equiv n-v \pmod 4$ and
the smallest $|b|$ (if any); $m_k$ is the minimum over the at most four residues. In an arithmetic
progression of step $g\mid 4$ each residue class modulo $4$ is a progression of step $4$, the whole
set, or empty, so the smallest $|a|$ in a class is found in $O(1)$. At the root $R_k$ is the full
parity class $\{-(n-k),\dots,n-k\}$. For odd $k$ both $C_k$ and
$C_{n-k}$ are odd and $a=b=1$ is feasible ($2\equiv n \pmod 4$), giving $2$. For even $k$ both are
even and $a+b\equiv 2 \pmod 4$ excludes $(0,0)$, giving $4$. The middle lag is odd, giving $1$. With
$n=4m+2$: $\LB=2m+4m+1=3n/2-2$.
\end{proof}

\section{Propagation and symmetry}
\label{app:prop}

This appendix defines the lag budgets and proves the budget propagation rule of
\cref{sec:branching}, condition (ii) of \cref{thm:framework}, and the symmetry reduction of
\cref{sec:bounds}, condition (iii).

The constraints attached to a node with threshold $T$ are the \emph{lag budgets}
\begin{equation}
  C_k(s)^2\le b_k,\qquad b_k=T-\LB_{-k}(\text{node}),\qquad k=1,\dots,n-1,
  \label{eq:budget}
\end{equation}
where $\LB_{-k}$ is a valid lower bound on $\sum_{j\ne k}C_j^2$ over the node's completions,
obtained from the bound of \cref{thm:pair} without lag $k$: for a lag $k$ paired with $n-k$ it is
$\LB-m_k+\mu_{n-k}^2$, and for the middle lag $\LB-\mu_k^2$. Every completion with $\E\le T$
satisfies them. The budget $b_k$ leaves the set $A_k=\{v\in R_k: v^2\le b_k\}$ of values that
$C_k$ can still take in a completion with $\E\le T$. If $A_k$ is the single value $\lo_k$, every
free chain of lag $k$ must attain its own minimum. A free chain with one fixed end attains its
minimum only when all of its products are $-1$, and so does a free chain with two fixed ends whose
minimum is $-e$; in both cases the free spins of the chain are determined uniquely from a fixed
end and are fixed. A free chain with two fixed ends and minimum $-(e-2)$, or one without a fixed
end, has several minimizing assignments, and nothing is forced. Symmetrically for $A_k=\{\hi_k\}$ with all products
$+1$. Whether $A_k$ is a single value is decided from $\lo_k$, $g_k$ and $b_k$ alone.

\begin{theorem}[budget propagation and infeasibility]\label{thm:prop}
Let $N$ be a node with threshold $T$, and let $\mu_k$, $m_k$ and $\LB$ be as in
\cref{thm:perlag,thm:pair}. For a lag $k$ paired with $k'=n-k$ set $\LB_{-k}=\LB-m_k+\mu_{k'}^2$,
and for the middle lag $\LB_{-k}=\LB-\mu_k^2$; set $b_k=T-\LB_{-k}$ and
$A_k=\{v\in R_k: v^2\le b_k\}$, with $R_k=\{\lo_k,\lo_k+g_k,\dots,\hi_k\}$.
\begin{enumerate}
\item[(1)] Every completion with $\E(s)\le T$ has $C_k(s)\in A_k$ for every $k$.
\item[(2)] If some $A_k$ is empty, $N$ has no completion with $\E\le T$.
\item[(3)] If $A_k=\{\lo_k\}$, every completion with $\E\le T$ has all free lag-$k$ products equal to
$-1$ in every free chain of type (ii), and in every free chain of type (i) whose minimum is $-e$;
the free spins of these chains are determined uniquely from their fixed ends. Symmetrically for
$A_k=\{\hi_k\}$ with all products $+1$. Fixing these spins removes no completion with $\E\le T$.
\item[(4)] $A_k=\{\lo_k\}$ holds iff $\lo_k^2\le b_k$ and ($\lo_k=\hi_k$ or $(\lo_k+g_k)^2>b_k$);
symmetrically for $\hi_k$.
\item[(5)] Repeating (1)--(3) until no spin changes terminates after at most $|U|$ rounds. If a
forcing assigns a value to a spin already fixed to the other value, or if $\LB$ exceeds $T$, $N$ has
no completion with $\E\le T$.
\end{enumerate}
Hence the rule ``force by (3) until no spin changes; report infeasible on (2), on a contradiction
or on $\LB>T$'' satisfies condition (ii).
\end{theorem}
\begin{remark}
The condition in (4) matters. A rule of the form ``$(|c_k|-f_k+2)^2>b_k$ and $c_k\ne0$'' on the
plain range $c_k\pm f_k$, with $f_k$ the number of free lag-$k$ products, without the condition
$|c_k|\ge f_k$ forces spins wrongly. For example, at $n=13$ take the node with
$s_0=s_2=s_3=s_5=s_9=+1$ fixed and the other eight spins free (0-based positions) and $T=14$.
For lag $1$, $c_1=1$, $f_1=11$ and $b_1=8$, so $(|c_1|-f_1+2)^2=64>b_1$ and the rule would force
all free lag-$1$ products to $-1$, that is, $C_1=-10$; but $R_1=\{-10,-8,\dots,12\}$,
$A_1=\{-2,0,2\}$, and the completion $(+,+,+,+,+,+,-,-,+,+,-,+,-)$ has $\E=14\le T$ and $C_1=2$.
All $R_j$ are recomputed after the first lag that forces anything, so every forcing uses the sets
of the current assignment.
\end{remark}

\begin{proof}[Proof of \cref{thm:prop}]
$\LB_{-k}$ is a valid bound on $\sum_{j\ne k}C_j^2$: dropping the pair term $m_k$ from $\LB$ and
re-adding $\mu_{k'}^2\le C_{k'}^2$ keeps a lower bound on the sum without lag $k$. (1) For a
completion with $\E\le T$, $C_k^2=\E-\sum_{j\ne k}C_j^2\le T-\LB_{-k}=b_k$, and $C_k\in R_k$ by
\cref{thm:perlag}. (2) is immediate from (1). (3) If $A_k=\{\lo_k\}$ then $C_k=\lo_k$, which is
$c_k$ plus the sum of the free chain minima, so the free chain sums add up to the sum of their
minima, and each free chain attains its own minimum. For a type-(ii) free chain the minimum $-e$
is attained only when all $e$ products are $-1$, which fixes its free spins from the fixed end $u$:
walking along the chain away from $u$, the $i$-th free spin is $(-1)^i s_u$. For a type-(i) free
chain with minimum $-e$ (parity allowing), likewise, and the walk from $u$ ends consistently at
the other fixed end $w$ because $(-1)^e=s_us_w$.
For a type-(i) free chain with minimum $-(e-2)$, or a type-(iii) free chain, the minimizing assignment
is not unique and nothing is forced. All forced values are necessary for $\E\le T$. (4) If
$\lo_k+g_k\le0$ then $(\lo_k+g_k)^2<\lo_k^2\le b_k$ and $A_k$ has two elements; if $\lo_k+g_k>0$,
every element above $\lo_k+g_k$ has a larger square than $(\lo_k+g_k)^2$, so $A_k=\{\lo_k\}$ iff
$(\lo_k+g_k)^2>b_k$. (5) After forcing, the sets $R_j$ shrink or stay (\cref{thm:perlag}), so
recomputation is valid; each round fixes at least one spin. Two necessary conditions for
$\E\le T$ that contradict each other, or $\LB>T$ (\cref{lem:lattice}(c)), show that no completion
has $\E\le T$.
\end{proof}

The symmetry rule of \cref{sec:bounds} rests on the following theorem. Negation and alternation
bring any sequence into the set $K=\{s: s_1=s_2=+1\}$, and the reversal rule compares a node's
prefix with the first entries of the element of $K$ obtained from the reversed sequence, which
depend only on the suffix.

\begin{theorem}[symmetry reduction]\label{thm:sym}
Let $\eta$ (negation) be $s_i\mapsto -s_i$, $\alpha$ (alternation) $s_i\mapsto(-1)^{i-1}s_i$ and
$\rho$ (reversal) $s_i\mapsto s_{n+1-i}$. They generate a group $G$ of order $8$ on which $\E$ is
constant. Let $K=\{s: s_1=s_2=+1\}$.
\begin{enumerate}
\item[(a)] Every orbit meets $K$: for each $s$ exactly one element $\kappa(s)$ of
$\{\eta^a\alpha^bs\}$ lies in $K$, and the orbit meets $K$ in exactly $\{\kappa(s),\kappa(\rho s)\}$.
Hence fixing $s_1=s_2=+1$ satisfies condition (iii).
\item[(b)] (Reversal rule.) Fix a total order on $\{+1,-1\}$ and let
$K'=\{s\in K: s\le_{\mathrm{lex}}\kappa(\rho s)\}$; $K'$ meets every orbit. Let a node have
$s_1=s_2=+1$, prefix $\pi=(s_1,\dots,s_m)$ and suffix $\tau=(s_{n-m+1},\dots,s_n)$ fixed, $m\ge2$.
The first $m$ entries $\pi'$ of $\kappa(\rho s)$ depend only on $\tau$. If $\pi'\lex\pi$ (decided
within the first $m$ entries), no completion of the node lies in $K'$; discarding such nodes
satisfies (iii) with $K'$ in place of $K$.
\end{enumerate}
\end{theorem}

\begin{proof}[Proof of \cref{thm:sym}]
$\eta$ leaves every product unchanged; $\alpha$ multiplies each lag-$k$ product by $(-1)^k$, so
$C_k\mapsto(-1)^kC_k$; $\rho$ reindexes, $C_k(\rho s)=\sum_i s_{n+1-i}s_{n+1-i-k}=C_k(s)$. All
three are involutions, $\eta$ commutes with $\alpha$ and $\rho$, and $\rho\alpha=\alpha\rho$ for
odd $n$, $\rho\alpha=\eta\alpha\rho$ for even $n$; the elements $\eta^a\alpha^b\rho^c$ form a group
of order $8$ \citep[App.~A, eqs.~(A.1)--(A.7)]{packebusch2016labs}. (a) $\eta$ flips $s_1$ and
$\alpha$ does not, so $a=[s_1=-1]$ makes $s_1=+1$; then $\alpha$ flips $s_2$ and keeps $s_1$, so
$b=[(-1)^as_2=-1]$, applied to the second entry after the negation, makes $s_2=+1$; the choice is
unique. The orbit is the union of the two cosets of
$\{\eta^a\alpha^b\}$ through $s$ and $\rho s$, each meeting $K$ once. (b) The subgroup
$\{\eta^a\alpha^b\}$ is normal in $G$ ($\rho\eta=\eta\rho$, and $\rho\alpha=\alpha\rho$ or
$\eta\alpha\rho$), so $\rho$ maps the coset $\{\eta^a\alpha^bs\}$ onto $\{\eta^a\alpha^b\rho s\}$
and $\kappa(\rho\kappa(s))=\kappa(\rho s)$. Hence $c=\lexmin(\kappa(s),\kappa(\rho s))$ satisfies
$c\le_{\mathrm{lex}}\kappa(\rho c)$, i.e.\ $c\in K'$, and $K'$ meets every orbit.
$\kappa(\rho s)_i=\varepsilon_i s_{n+1-i}$ with $\varepsilon_i=s_n(-1)^{(i-1)t}$,
$t=[s_ns_{n-1}=-1]$, so the sign pattern depends only on $s_n,s_{n-1}$ and $\pi'$ only on $\tau$.
For a completion $c$ of the node with $c\in K'$ we would need $c\le_{\mathrm{lex}}\kappa(\rho c)$,
whose first $m$ entries are $\pi$ and $\pi'$; $\pi'\lex\pi$ decided within $m$ entries contradicts
this. So no completion lies in $K'$.
\end{proof}

\section{Conflict values}
\label{app:conflict}

This appendix defines the LABS conflict values of \cref{sec:branching} formally and proves the two
propositions stated there. Neither is used in the proof of correctness.

\begin{definition}[conflict values for LABS]\label{def:conflict}
Let a node $(\Phi,U)$ have the lag budgets $b_k$ of \cref{eq:budget}, and let its samples,
completed to full sequences, be the distinct sequences $s^\ell$ with counts $\omega_\ell$,
$q=\sum_\ell\omega_\ell$. The violation matrix of
QCBB eq.~(5) and the correspondence matrix of QCBB eq.~(7) are
\begin{equation}
  V_{k\ell}=[\,C_k(s^\ell)^2>b_k\,],\qquad
  P_{ki}=[\,i-k\in\Phi\ \text{or}\ i+k\in\Phi\,]\ \ (i\in U),
  \label{eq:conflict}
\end{equation}
and $\nu$ and $\chi=\nu P$ are QCBB eqs.~(6) and (8) unchanged, as in
\cref{eq:conflict-main}. $P_{ki}=1$ iff spin $i$ has a
fixed partner at lag $k$, that is, iff $i$ enters the linear
part of $C_k$ in \cref{eq:sub} with coefficient $h_{ki}$ (which may cancel to $0$ when both
partners are fixed). The branching spin is $i^*=\arg\max\chi_i$ over the candidate set, QCBB
eq.~(9), with ties broken by the smallest key $(d(i),\text{side})$, $d(i)=\min(i,n-1-i)$
(0-based), where $\text{side}=i$ if the number of fixed spins of the node is even and
$\text{side}=-i$ if it is odd, so that the left spin wins for an even and the right spin for an
odd fixed count.
\end{definition}

With a prefix and a suffix of $m$ fixed spins each, $P_{\cdot i}$ has $2m$ ones for every free
spin except those near the middle, where the lag ranges $[i-m,i-1]$ and $[n-m+1-i,n-i]$ overlap
and the count drops (for $n=30$, $m=5$: $\chi=10$ for 16 spins and $8,6,6,8$ for the four middle
ones with $\nu\equiv1$). So eq.~(9) over all free spins with $\nu\equiv1$ selects one of the two
outermost free spins, and if the prefix and the suffix have the same length, the two outermost
free spins have equal conflict values for every $\nu$ (\cref{prop:frontier}). On every such node
the restriction to the two outermost spins is therefore eq.~(9) itself, evaluated on two spins
instead of $F$. Eq.~(9) over
all free spins can differ from it only at nodes where propagation has fixed interior spins, or,
with sample $\nu$, where the samples move the maximum inward. With sample $\nu$, long lags have
few free products, tight budgets and high $\nu_k$, so their weight falls on spins near the ends.
Once $\UB=E^*$, the threshold no longer changes, and whether a node is pruned depends only on the
node; the samples then act only through the choice between the two outermost free spins
(\cref{prop:samples}). The two propositions below make these statements exact.

\begin{proposition}[eq.~(9) on nodes with a fixed prefix and suffix]\label{prop:frontier}
Let a node have the fixed set $\{0,\dots,a-1\}\cup\{b+1,\dots,n-1\}$ (0-based), free block
$U=\{a,\dots,b\}$, $r=n-1-b$, $a+r\ge1$, and let ties go to the smallest key $(d(i),\text{side})$
with $d(i)=\min(i,n-1-i)$.
\begin{enumerate}
\item[(a)] With $\nu\equiv1$, eq.~(9) over all free spins selects $a$ or $b$.
\item[(b)] If $a=r$, then $\chi_a=\chi_b$ for every $\nu\ge0$.
\end{enumerate}
\end{proposition}

\begin{proof}[Proof of \cref{prop:frontier}]
(a) For $i\in U$, lag $k$ gives $i$ a fixed partner on the left iff $k\in\mathcal{L}(i)$ and on
the right iff $k\in\mathcal{R}(i)$, where
\begin{equation*}
  \mathcal{L}(i)=\{i-a+1,\dots,i\},\qquad \mathcal{R}(i)=\{b+1-i,\dots,n-1-i\}
\end{equation*}
have $a$ and $r$ elements, all in $\{1,\dots,n-1\}$. With $\nu\equiv1$,
$\chi_i=|\mathcal{L}(i)\cup\mathcal{R}(i)|=a+r-O(i)$ with
\begin{equation*}
  O(i)=|\mathcal{L}(i)\cap\mathcal{R}(i)|=\max\bigl(0,\min(i,n-1-i)-\max(i-a+1,b+1-i)+1\bigr).
\end{equation*}
Let $a<i<b$ (so $F=b-a+1\ge3$) and first $i\le n-1-i$. Then $a<i\le(n-1)/2$, so
$d(a)=a<i=d(i)$: $a$ has the strictly smaller key. Further $O(a)=\max(0,a-F+1)$, since
$\min(a,n-1-a)=a$ and $\max(1,b+1-a)=F$, while $\min(i,n-1-i)=i\ge a+1$ and
$i-a+1,\,b+1-i\le F-1$ give
\begin{equation*}
  O(i)\ge\max(0,a-F+3)\ge O(a),
\end{equation*}
so $\chi_i\le\chi_a$. If
$\chi_i$ is not maximal, $i$ is not selected; if it is, $a$ is maximal too and wins the tie. For
$i>n-1-i$ the reflection $j\mapsto n-1-j$ swaps prefix and suffix, maps $b$ to the left end and
keeps $\chi$ and $d$, so the same argument applies with $b$. (b) If $a=r$ the reflection maps the
fixed set onto itself and $a$ onto $b$, so $P_{ka}=P_{kb}$ for every $k$ and
$(\nu P)_a=(\nu P)_b$.
\end{proof}

\begin{proposition}[what the samples change in a proof]\label{prop:samples}
Consider a run without a node or time limit, with the two outermost free spins as branching
candidates, and let $t^*$ be
the first time at which $\UB=E^*$.
\begin{enumerate}
\item[(a)] After $t^*$ the threshold is $T=E^*-\Delta$ and does not change.
\item[(b)] Whether a node evaluated after $t^*$ is dropped (reversal rule, propagation or bound)
depends only on the node and $T$, not on any sample.
\item[(c)] Every node that is not dropped and has a free spin is sampled and branched; for every
node other than the master, its samples choose the branching spin among the two outermost free
spins,
and every child whose key is at most $T$ is evaluated before the run ends.
\end{enumerate}
Hence the samples change the set of evaluated nodes in two ways only: through $t^*$, since every
node evaluated before $t^*$ is judged at a larger threshold, and through the choice of the
outermost spin at each branched node. At a node whose fixed spins form a prefix and a suffix of
equal length, with no fixed interior spin, this choice does not depend on the samples.
\end{proposition}

\begin{proof}[Proof of \cref{prop:samples}]
(a) A leaf or a sample lowers $\UB$ only if its energy is below $\UB=E^*$, which is impossible.
(b) The reversal rule, the propagation rule and the bound use only the node and $T$
(\cref{thm:sym,thm:prop,thm:pair}). (c) The first part is the definition of the node loop
(\cref{alg:main}). The run ends only when the lowest key in the queue exceeds $T$, so every child
that entered the queue with a key at most $T$ is evaluated first. Given $t^*$ and these choices,
(a)--(c) therefore determine the evaluated set. The last statement is
\cref{prop:frontier}(b): the two outermost free spins have equal conflict values for every
$\nu$, and
the tie rule decides.
\end{proof}

\section{Repeated runs}
\label{app:seeds}

QOBLIB recommends repeated independent runs of a stochastic method, so we repeated the QAOA
runs at $p=12$ of \cref{tab:backends} with the seeds $1$--$5$, with the settings of
\cref{sec:setup} and one run at a time. All $145$ runs, labs002--labs030 with five seeds each,
proved the QOBLIB optimum, and their sequences, summaries and incumbent time series form our
QOBLIB submission. For $n\le15$ every run takes at most $6$ nodes and $0.04$\,s, and
\cref{tab:seeds} gives $n=16$--$30$. Since the samples change the search tree only through the
time at which the optimum is first found and through the choice of the branching spin
(\cref{prop:samples}), the seed matters little here: up to $n=24$ every seed evaluates the same
tree, and above it the node counts of the seeds differ by at most $0.2\%$, with one exception. At
$n=25$ the run with seed~$4$ found the optimum only at the end of the search, after $14.8$\,s
instead of about $2.5$\,s, and evaluated $13{,}756$ nodes instead of about $6{,}140$, which is the
effect of a late incumbent described in \cref{sec:samples}; its first call returned $44$ instead
of $E^*=36$, the only one of the $5\times11$ first calls at $n=16$--$26$ to miss $E^*$ (last
column of \cref{tab:seeds}). Seed~$1$ evaluates exactly the trees
of \cref{tab:backends}, so the runtimes of that run and of these repeats differ only by the
variation of the measured time; the range of the five runtimes is within $20\%$ of its lower end
at $n\le21$, where the runs take less than half a second, and within $9\%$ at $n=22$--$30$
($n=25$ excepted).

\begin{table}[tbp]
\centering
\small\setlength{\tabcolsep}{4pt}
\begin{adjustbox}{max width=\textwidth}
\begin{tabular}{rr rr rr rr r r}
\toprule
 & & \multicolumn{2}{c}{nodes} & \multicolumn{2}{c}{runtime (s)} & \multicolumn{2}{c}{time to $E^*$ (s)} & PDI & $E^*$ at\\
\cmidrule(lr){3-4}\cmidrule(lr){5-6}\cmidrule(lr){7-8}
$n$ & $E^*$ & median & range & median & range & median & range & (s) & first call\\
\midrule
16 & 24 & 61 & -- & 0.05 & 0.05--0.06 & 0.05 & 0.05--0.06 & 0.05 & 5/5\\
17 & 32 & 319 & -- & 0.07 & 0.07--0.07 & 0.06 & 0.06--0.06 & 0.06 & 5/5\\
18 & 25 & 1 & -- & 0.04 & 0.04--0.05 & 0.04 & 0.04--0.05 & 0.04 & 5/5\\
19 & 29 & 411 & -- & 0.14 & 0.14--0.14 & 0.09 & 0.08--0.09 & 0.10 & 5/5\\
20 & 26 & 328 & -- & 0.22 & 0.22--0.23 & 0.12 & 0.11--0.12 & 0.16 & 5/5\\
21 & 26 & 381 & -- & 0.30 & 0.30--0.34 & 0.18 & 0.17--0.21 & 0.22 & 5/5\\
22 & 39 & 1{,}777 & -- & 1.11 & 1.06--1.14 & 0.32 & 0.28--0.34 & 0.48 & 5/5\\
23 & 47 & 14{,}604 & -- & 8.71 & 8.68--8.99 & 0.60 & 0.56--0.68 & 3.25 & 5/5\\
24 & 36 & 4{,}643 & -- & 4.67 & 4.53--4.83 & 1.23 & 1.09--1.28 & 2.44 & 5/5\\
25 & 36 & 6{,}141 & 6{,}139--13{,}756 & 7.40 & 7.25--14.84 & 2.50 & 2.37--14.84 & 4.49 & 4/5\\
26 & 45 & 7{,}080 & 7{,}076--7{,}084 & 10.99 & 10.39--11.27 & 5.28 & 4.63--5.37 & 6.10 & 5/5\\
27 & 37 & 12{,}433 & 12{,}427--12{,}435 & 13.27 & 13.22--13.61 & 0.04 & 0.03--0.04 & 4.03 & 5/5\\
28 & 50 & 134{,}199 & 134{,}031--134{,}205 & 131.87 & 130.69--134.95 & 24.22 & 24.01--24.85 & 48.02 & --\\
29 & 62 & 443{,}142 & 443{,}126--443{,}714 & 449.70 & 444.62--464.19 & 0.04 & 0.04--0.04 & 159.73 & 5/5\\
30 & 59 & 182{,}730 & 182{,}727--182{,}737 & 232.54 & 232.35--240.00 & 106.50 & 106.36--110.71 & 71.78 & --\\
\bottomrule
\end{tabular}
\end{adjustbox}
\caption{Five runs per instance of QAOA at $p=12$ with the
seeds $1$--$5$ (settings of \cref{sec:setup}, $1{,}000$ shots per node; measured). Every run
proves the QOBLIB optimum $E^*$. The range of the node counts is omitted (--) where all five
runs evaluate the same tree. The runtime is the time to the proof and the time to $E^*$ the time
until the incumbent first reaches $E^*$; both are given as the median and the range over the
five runs, and the PDI as the mean. The last column counts the seeds whose first sampler call,
on the master (for $n=27$ and $29$ on its skew-symmetric restriction, \cref{sec:setup}), returned
a sequence of energy $E^*$; at $n=28$ and $30$ the master is not sampled (--).}
\label{tab:seeds}
\end{table}

\end{document}